\documentclass[preprints,review,accept,pdftex,moreauthors]{Definitions/mdpi} 

\firstpage{1} 
\pubvolume{1}
\issuenum{1}
\articlenumber{0}
\pubyear{2023}
\copyrightyear{2023}
\datereceived{ } 
\daterevised{ } 
\dateaccepted{ } 
\datepublished{ } 
\hreflink{https://doi.org/} 
\pdfoutput=1 

\usepackage{graphicx}
\usepackage{subcaption}
\usepackage{txfonts}
\usepackage{isotope}

\newcommand{\msun}{$\rm M_{\odot}$}
\newcommand{\p}{$p$}

\newcommand{\g}{$\gamma$}
\newcommand{\A}{$\alpha$}

\newcommand{\fz}{$\rm F_0$}
\newcommand{\fg}{$\rm F_{\gamma}$}
\newcommand{\fo}{$\rm F_{O16}$}

\graphicspath{{./}{figures/}}

\newcommand\aj{AJ}                   
\newcommand\araa{ARA\&A}             
\newcommand\fcp{Fundamentals Cosmic Phys.}  
\newcommand\nar{New~Astron.~Rev.}    
\newcommand\nphysa{Nuclear Phys.~A}  
\newcommand\prc{Phys. Rev.~C}        
\newcommand\prd{Phys. Rev.~D}        
\newcommand\prl{Phys. Rev.~Lett.}    
\newcommand\physrep{Phys.~Rep.}      
\newcommand\ssr{Space Sci. Rev.}     
\newcommand\zap{Z.~Astrophys.}       

\Title{The physics of Core-Collapse Supernovae: explosion mechanism and explosive nucleosynthesis}

\TitleCitation{The physics of CCSNe: explosion mechanism and explosive nucleosynthesis}

\Author{Luca Boccioli $^{1,\dagger}$*\orcidA{}, Lorenzo Roberti $^{2}$ $^{3,}$ $^{4,\dagger}$*\orcidB{}}

\AuthorNames{Luca Boccioli, Lorenzo Roberti}

\AuthorCitation{Boccioli, L.; Roberti, L.}

\address{%
$^{1}$ \quad Department of Physics, University of California, Berkeley, CA 94720, USA \\
$^{2}$ \quad Konkoly Observatory, Research Centre for Astronomy and Earth Sciences, HUN-REN, Konkoly Thege Mikl\'{o}s \'{u}t 15-17, H-1121 Budapest, Hungary\\
$^{3}$ \quad CSFK, MTA Centre of Excellence, Budapest, Konkoly Thege Miklós út 15-17, H-1121, Hungary\\
$^{4}$ \quad INAF -- Osservatorio Astronomico di Roma Via Frascati 33, I-00040, Monteporzio Catone, Italy}
                
\corres{Correspondence: lbocciol@berkeley.com (L.B.); lorenzo.roberti@csfk.org (L.R.)}

\firstnote{These authors contributed equally to this work.}

\abstract{Recent developments in multi-dimensional simulations of core-collapse supernovae have considerably improved our understanding of this complex phenomenon. In addition to that, one-dimensional (1D) studies have been employed to study the explosion mechanism and its causal connection to the pre-collapse structure of the star, as well as to explore the vast parameter space of supernovae. Nonetheless, many uncertainties still affect the late stages of the evolution of massive stars, their collapse, and the subsequent shock propagation. In this review, we will briefly summarize the state-of-the-art of both 1D and 3D simulations and how they can be employed to study the evolution of massive stars, supernova explosions, and shock propagation, focusing on the uncertainties that affect each of these phases. Finally, we will illustrate the typical nucleosynthesis products that emerge from the explosion.}


\begin{document}

\section{Introduction}

Massive stars (M$\rm _{ZAMS}$ > 8--10 \msun) end their lives with the formation and subsequent collapse of a Fe core. Over the years, the multiple efforts of the stellar community have produced evolutionary codes able to accurately simulate the life of these objects \citep{weaver:78,Woosley2002_KEPLER_models,heger:10,ekstroem:08,ekstrom:12,chieffi:98,CL13,LC18,paxton:11,paxton:18}.
However, the calculation of the complete evolution of a massive star, from H burning up to the formation of the Fe core is still characterized by several physical and computational challenges \citep[e.g.][]{stancliffe:15,stancliffe:16,rizzuti:22,rizzuti:23}. 
Mixing and transport processes, magnetic fields, mass-loss, and nuclear reaction rates are only some of the many uncertainties that affect the evolution of massive stars. 

The subsequent phase, i.e. a core-collapse supernova (CCSN), gives rise to new sources of uncertainties. The wide range of densities ($10^3-10^{15}$ g cm$^{-3}$), temperatures ($1-100$ MeV), and electron fractions ($0.01-0.6$) that characterize a supernova pose an extremely challenging problem. To accurately model a CCSN, one needs a detailed description of the properties of matter at extremely high densities and temperatures, i.e., the equation of state (EOS) of nuclear matter, which is however still poorly constrained (e.g. \cite{Klahn2006_HIC_constraints_EOS,Hebeler2013_constraints_EOS,Zhang2018_constraints_on_EOS,Burgio2021_constraints_on_EOS,Raduta2021_EOS_review,Furusawa2023_EOS_review,Nattila2016_EOS_constraints,Radice2018_contraints_on_EOS_from_GW170817,PREX2021_nskin_208Pb,CREX2022_nskin_40Ca}). As a consequence of the high densities reached deep inside the proto-neutron star (PNS), general relativistic effects can play a non-negligible role \citep{Boccioli2021_STIR_GR}. Moreover, at such high densities, the mean free path of neutrinos is very small, and therefore they are trapped inside the PNS and are only able to escape once the density goes below $\sim 10^{11}-10^{12}$ g cm$^{-3}$. Since a huge amount of neutrinos is produced, and they are responsible for triggering the explosion, one needs an accurate description of the interactions and transport of those neutrinos within the star \citep{Mezzacappa2020_nu_transport_review,Richers2017_nu_transport_comparison,Fischer2017_Review_EOS_nu}. Finally, self-consistent, high-resolution simulations are required to capture multi-dimensional effects that play a crucial role in the explosion \citep{Muller2012_2D_GR,Takiwaki2012_original_3DnSNe,Lentz2015_3D,Janka2016_success_expl,Bruenn2016_expl_en,Takiwaki2016_3DnSNe_3D_explosions,OConnor2018_2D_M1,Muller2019_3Dcoconut,Burrows2020_3DFornax,Bugli2021_MHD_3D_SN,Nakamura2022_3DnSNe_binary_star_1987a}.

CCSNe are also responsible for creating half of the elements of the periodic table. Given the extreme and diverse thermodynamic conditions achieved in the explosion, several nucleosynthetic processes may occur, as, e.g., the explosive burning stages relative to the major hydrostatic fuels of the stars (Si, O, Ne, C, He), $\nu$p--process, (weak) $r$--process, or \g--process. Most of these are currently performed in post-process, i.e., the nuclear network calculations are computed on the output of hydrodynamic simulations, as opposed to being computed out on-the-fly, which is much more computationally expensive and unfeasible with current codes, although promising efforts are underway \citep[]{sandoval:21,rizzuti:23}. These nucleosynthesis calculations are therefore heavily dependent on the characteristics of the explosion, mainly the explosion energy and the mass cut, and on the uncertainties in nuclear reaction rates.

In this review, we will summarize the pre-supernova evolution and uncertainties in the stellar evolution. Then, we will review the current state-of-the-art of CCSNe, with a particular focus on the nuclear EOS and the explodability problem. Finally, we provide a general overview of explosive nucleosynthesis.

\section{The progenitors of CCSNe: massive stars and their evolution} \label{sec:evo}
The evolution of massive stars is governed by a sequence of nuclear burning stages, in which lighter nuclear species are progressively transformed into heavier ones. Each burning stage primarily takes place in the core of the star and then it shifts outward in mass (shell burning) as soon as the available fuel diminishes. Normally, shell burning takes place at higher temperatures and lower densities in comparison to central burning. In each successive burning phase, the main product from the previous phase becomes the primary fuel, until the formation of a dense core made predominantly of Fe-peak elements (iron core). The energy released through nuclear reactions plays a crucial role in counterbalancing the gravitational collapse, especially in the advanced stages of stellar evolution, where neutrino losses also come into play. In the following, we briefly discuss the key aspects of the evolution that lead massive stars to explode as a CCSN. 

\subsection{Pre-supernova evolution}

Six major fuels identify the principal burning stages in massive stars, , i.e., H, \isotope[4]{He}, \isotope[12]{C}, \isotope[20]{Ne}, \isotope[16]{O}, and \isotope[28]{Si} \citep{kippenhahn:90}.

In the core of massive stars, H fusion occurs through the CNO cycle, while the proton-proton (p-p) chain plays only a marginal role \citep{cameron:57,clayton:68}. The conversion of H in He lasts a few million years and scales inversely with the initial mass of the star. H burning in massive stars is characterized by the presence of a convective core which recedes in mass by decreasing the amount of available H. Once the H is depleted in the core, the star contracts until the He burning reactions, i.e., the $3\alpha$ reaction and \isotope[12]{C}($\alpha$,$\gamma$)\isotope[16]{O}, are activated, while the H burning shifts in a shell. 
In this phase, the radius expands, the surface cools down, and the star becomes a red supergiant (RSG). Moreover, mass-loss may critically increase and play a crucial role, since it may reduce the size of the H-depleted core (He core) and therefore modify the evolutionary path of the star. 
Core He burning occurs in a convective core as well and it is controlled by the $3\alpha$ reaction in the early stages, which produces \isotope[12]{C}. As \isotope[12]{C} increases, the \isotope[12]{C}($\alpha$,$\gamma$)\isotope[16]{O} reaction gradually becomes more and more efficient, until it takes over the $3\alpha$ reaction. The competition between these two nuclear processes controls the C/O ratio at the time of the He exhaustion in the core. 

At the end of central He burning, two main quantities govern the further evolution of the star: the \isotope[12]{C}/\isotope[16]{O} ratio and the mass of the CO core left by the He depletion. The CO core mass plays the same role the initial mass has in H burning. The \isotope[12]{C} abundance left by central He burning influences both the formation of the convective core during the central C burning phase and the behavior of the C burning shells during the central burning of Ne, O, and Si, contributing to determine the final compactness of the star and eventually its final fate (see Sect. \ref{sec:explodability}). Neutrino energy losses begin to become relevant at a temperature of the order of $\rm \sim800~ MK$, i.e., the typical temperature of C ignition, and enter into the total energy balance together with gravitational and nuclear energy. From this point on, the evolution of the star can be described by a sequence of progressively faster contraction, central and shell burning stages, that lead \isotope[12]{C} to be converted into \isotope[20]{Ne}, \isotope[20]{Ne} into \isotope[16]{O}, \isotope[16]{O} into \isotope[28]{Si}, and finally \isotope[28]{Si} into Fe-peak elements. 

It is however interesting to note that the processes that lead from the exhausted O core to the formation of the iron core deviate from the classical definition of nuclear burning. Instead, they can be interpreted as a re-adjustment of the chemical composition after the nuclear reactions have reached a (partial) equilibrium. For a detailed discussion of the latest stages of massive stars, we refer the reader to more specialized works \citep[e.g.][]{woosley:95,Woosley2002_KEPLER_models,nomoto:13,LC18}. 

\begin{figure}[!t]
\includegraphics[width=\columnwidth]{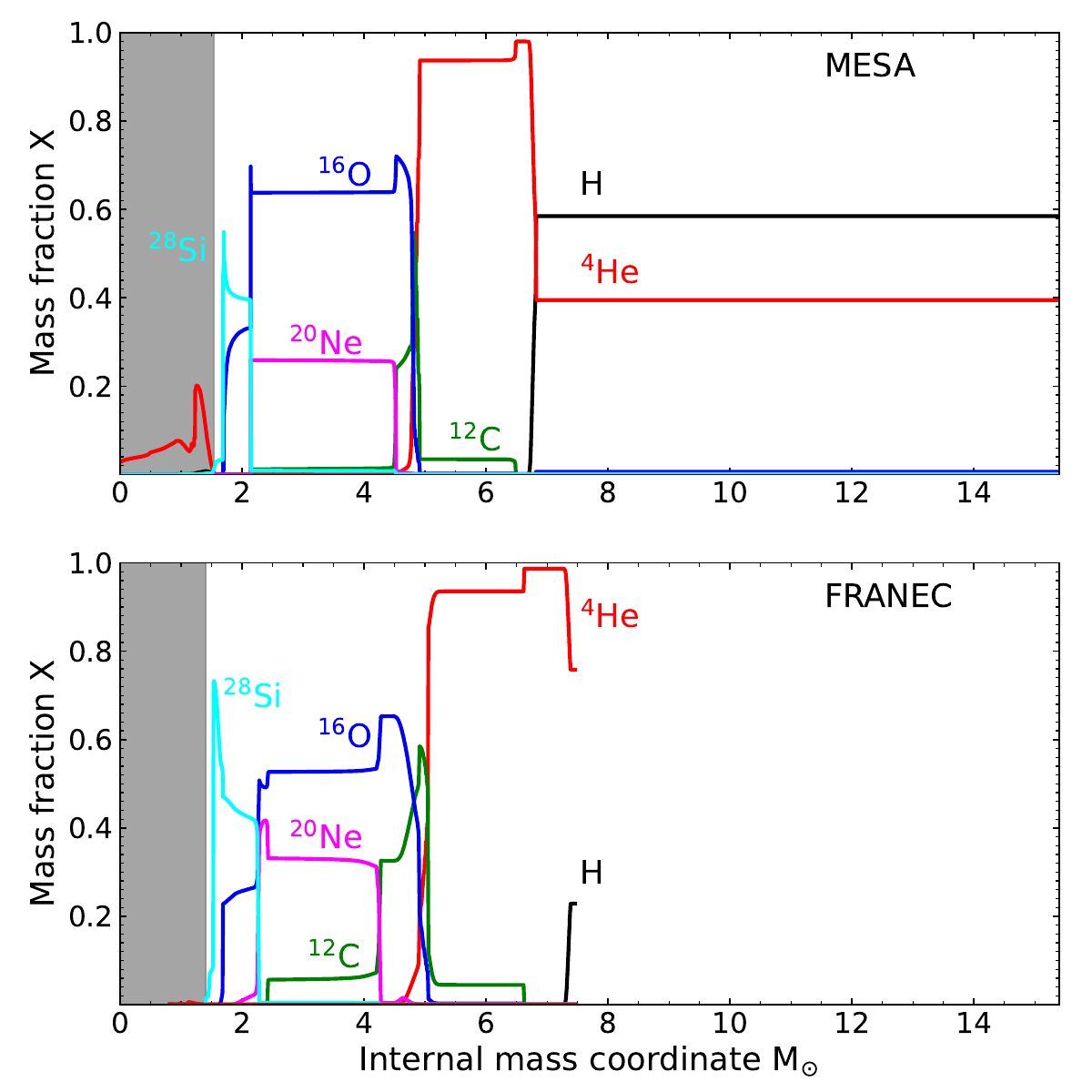}
\caption{Internal composition of a non-rotating 20 \msun\ star at solar metallicity at the core-collapse stage. The upper panel represents a model calculated with the MESA code. The lower panel represents a model calculated with the FRANEC code instead. The different colors mark the different abundances in mass fraction as a function of the internal mass coordinate, from the center up to the surface of each model. The grey shaded area represents the iron core. \label{fig:psn}}
\end{figure}  

\figurename~\ref{fig:psn} shows the distribution of the abundances of the major fuels in a non-rotating 20 \msun\ star at solar metallicity, calculated with two different stellar evolutionary codes, namely \verb|FRANEC| \citep{CL13,LC18,roberti:23b}, and \verb|MESA| \citep{paxton:10,paxton:15,ritter:18}. As discussed in the next section, different codes can produce significantly different results for the same set of initial conditions (e.g., mass, metallicity, rotation, magnetic field...). 

\subsection{Challenges and uncertainties in stellar evolution} \label{sec:uncertainties}

While significant progress has been made in modeling the evolution of massive stars, several challenges and uncertainties still persist. In the following, we mention and briefly discuss the main open problems affecting the calculation of CCSN models. Note that this list does not claim to be exhaustive and only provides a starting point for the reader to frame the open problems in the modeling of stellar evolution.

\subsubsection{Mass-loss}
Accurate modeling of mass-loss in massive stars remains a challenge. The mechanisms responsible for mass-loss, such as radiation-driven winds and eruptions, are not fully understood and they can have a critical impact on the evolution, the nucleosynthesis, and on the final fate of massive stars \citep[e.g.][]{langer:12,smith:14}. The mass-loss prescriptions in stellar models are often empirical and rely on arbitrary parameters; moreover, their dependency on the metallicity of the star is still unclear.
    
\subsubsection{Mixing and transport processes}
The most significant source of uncertainty in stellar models is undoubtedly represented by the treatment of convection, which is intrinsically a multi-dimensional mixing and transport process. It is impossible to determine the convective transport, the extension of the convective zones, and the velocity of the eddies of material by first principles only. In order to include its (inherently multi-dimensional) effects in 1D, spherically symmetric codes, it is commonly modeled by means of the mixing length theory (MLT). Broadly speaking, the larger the magnitude of the mixing length, the stronger the convection. The mixing length is usually expressed as:
\begin{equation}
    \label{eq:MLT}
    \Lambda_{\rm MLT} = \alpha_{\rm MLT} \frac{P}{\rho g},
\end{equation}
where $\alpha_{\rm MLT}$ is a parameter of order $\sim \mathcal{O}(1)$, and $P$, $\rho$, and $g$ are the local pressure, density, and gravitational acceleration, respectively. Typically, the value of $\alpha_{\rm MLT}$ is calibrated to reproduce the properties of the Sun \citep{chandrasekhar:39,BohmVitense1958,kippenhahn:90}. There is a huge amount of literature on this topic and definitions of the mixing length beyond the simple expression given in eq. \eqref{eq:MLT} have been proposed, as well as models beyond MLT calibrated on multi-dimensional simulations \citep{Arnett2015_321D}. However, this goes beyond the scope of this article, and we point the reader to dedicated reviews on this topic \citep{Joyce2023_MLT_review}.
    
Another intrinsically multi-dimensional phenomenon in stars is rotation, which is responsible for the transport of angular momentum and chemical species within the star. Rotation is usually parameterized in 1D and calibrated to reproduce the observed properties of a sample of nearby rotating stars, such as the surface chemical enrichment and the equatorial velocity \citep{meynet:97,maeder:00,CL13,LC18}. 
    
These two processes play a significant role in determining the distribution of elements within the star, impacting subsequent nucleosynthesis, the composition of the stellar envelope, and eventually even enhancing the mass lost from the surface of the star. In particular, they highlight the need to develop multi-dimensional stellar codes that can adequately treat the transport mechanisms inside a star.

\subsubsection{Magnetic fields}
The role of magnetic fields in massive star evolution is not well-constrained. Much like mass-loss, convection, and rotation, it depends on arbitrary parameters. Modeling stellar magnetic fields involves solving magneto-hydrodynamic (MHD) equations, which describe the interactions between magnetic fields and fluid motion. Accurate numerical solutions are computationally intensive and require advanced algorithms \cite{spruit:02}. Therefore, as it is common practice in studying the evolution of massive stars, we will ignore the effects of magnetic fields in the remainder of this paper.

\subsubsection{Nuclear reaction rates}
Nuclear reaction rates are responsible for the energy generation and the nucleosynthesis of all the nuclear species in every layer of the star. Large uncertainties affect these reaction rates, which propagate, in turn, on the calculation of massive star models \citep[e.g.][]{imbriani:01,chieffi:21}. More precise experimental data and improved theoretical models are needed to better constrain these rates. 

\subsubsection{Binary nature of massive stars}
The majority of massive stars are in binary systems, and interactions with a companion star can have a significant impact on their evolution \citep{langer:12,laplace:21,brinkman:21,brinkman:23,ercolino:23}. Understanding the effects of binary interactions, such as mass transfer, common envelope evolution, and mergers, is a current challenge that in recent years is becoming more and more approachable with modern tools \citep{chen:24}. For the remainder of this paper, we will ignore binary interactions and focus solely on the evolution of single stars.

\subsubsection{Core-collapse supernova mechanism}
The core-collapse supernova mechanism, responsible for the explosive death of massive stars, is still not fully understood. The details of the shock revival, neutrino interactions, and explosion dynamics are complex and require advanced and multi-dimensional numerical simulations. We will discuss this topic in detail in Sect. \ref{sec:SN_engine}. 

\subsubsection{Stellar codes} \label{sec:codes}
Performing accurate and computationally intensive simulations of massive star evolution, including hydrodynamics, radiation transport, and nuclear reactions, remains a challenge. Moreover, the adoption of different criteria for mass-loss, convection, rotation, magnetic fields, as well as of different physics inputs and different numerical methods, leads inevitably to a degeneration of solutions for the calculation of a model using the same initial conditions \citep[see, e.g.,][and  \figurename~\ref{fig:psn}]{jones:15,bruenn:23}. Regarding this last point, one key aspect is the nuclear network adopted, namely, the list of all the isotopes explicitly included in the calculation and of all the reactions that link them with each other. Recent studies \citep[e.g.,][]{farmer:16} showed in fact that large nuclear networks are required especially in the latest stages of the evolution of massive stars, where multiple reactions can contribute to the energy generation that sustains the structure and regulate the behavior of burning shells. Underestimating the number of nuclear species in the nuclear network may substantially decrease the computation time and facilitate the convergence of models, but it may also lead to significantly different results relative to the calculations made using larger nuclear networks. For this reason, increasing computational capabilities is an ongoing effort to produce more refined and accurate simulations. Moreover, as mentioned above, a multi-dimensional approach is required, but still far from being realized.


\section{Core-collapse supernova theory}
\label{sec:SN_engine}
The first theoretical models of supernovae date back to the late fifties \citep{BBFH_1957,Hoyle1960_SN_Nucleosynthesis}. In these seminal papers, it was postulated that a supernova explosion could be powered by thermonuclear burning of the material right outside the core. This was later shown to be incorrect by the first numerical simulations, \citep{Colgate_White1966,Arnett1966}, which instead recognized the crucial role of neutrinos as being the primary cause of the explosion. 

In these simulations, neutrinos emitted during the collapse would deposit energy behind the shock, energizing the explosion right after core bounce. Later simulations \citep{Bethe_Wilson1985} that employed more detailed microphysics, showed that this so-called "prompt neutrino-driven mechanism" could not provide enough energy to power an explosion. Instead, they found that a "delayed neutrino-heating", after the initial expansion and a brief stalling of the shock, could be responsible for reviving the shock at later times and launching the supernova.

Broadly speaking, this is a well-accepted explosion mechanism even today. However, significant theoretical and computational efforts have shown that a supernova is an extremely complicated interplay between microphysics \citep{Lattimer1991_LS, Chabanat1997_SLy4, Shen1998_original_HShen, Hempel2010_HS_RMF, Steiner2013_SFHo, Dutra2014_Skyrme_params, Schneider2017_SROEOS,Bruenn1985, Mezzacappa1993a_infall, Mezzacappa1993b_method, Mezzacappa1993c_nu_e_scat, Thompson2000_mutau_therm, Horowitz2002, BRT2006, Fischer2017_Review_EOS_nu,Herwig2000}, radiation transport, and magneto-hydrodynamics \citep{Mezzacappa1999_nutransport,Liebendorfer2004_GR_nutransport,Mezzacappa2020_nu_transport_review,Glas2019_comparison_nu_transport_AA,Pan2019_nu_transport_2D_comparison,Richers2017_nu_transport_comparison}, where multi-dimensional effects play a crucial role \citep{Radice2016,Radice2018_turbulence,Muller2012_2D_GR,Takiwaki2012_original_3DnSNe,Lentz2015_3D,Janka2016_success_expl,Bruenn2016_expl_en,Takiwaki2016_3DnSNe_3D_explosions,OConnor2018_2D_M1,Muller2019_3Dcoconut,Burrows2020_3DFornax,Bugli2021_MHD_3D_SN,Nakamura2022_3DnSNe_binary_star_1987a}.

Nonetheless, it is of significant pedagogical value to illustrate the physical mechanisms behind a simplified model of the explosion.

\subsection{The collapse phase}
After a massive star has built up its iron core, nuclear reactions turn off, since iron has the largest binding energy among all nuclear species. This removes the pressure support generated by nuclear energy that prevents the core from collapsing. Therefore, gravity becomes the dominant force and this triggers core-collapse. At this point, electrons are degenerate, and neutrino emission is the main process responsible for the cooling of the iron core. As the core collapses, densities and temperatures increase, which lifts electron degeneracy, and therefore facilitates electron capture on free protons and nuclei, which deleptonizes the core. As a consequence, the electron fraction goes from $\sim 0.42-0.45$ down to $\sim 0.3$. Neutrino emission tends to decrease the entropy, whereas electron captures on nuclei and photodissociation tend to increase it \citep{Branch&Wheeler2017_SN_explosions}. The net result is that entropy is roughly constant, and the collapse is therefore adiabatic. Together with the extremely low (practically zero) pressure which fails to balance gravity, this causes the collapse to be homologous, i.e. $v \propto r$, where $v$ and $r$ are the velocity and radius, respectively. This means that there is some radius $r_{\rm s}$ (the sonic point) at which the speed of the collapse is greater than the sound speed. Beyond that point, pressure waves cannot propagate fast enough to re-equilibrate the changing pressure gradients, and therefore matter is in approximate free fall.

The central density increases extremely rapidly and at some point the neutrinos that are being emitted become trapped, at density $\rho_{\rm trap} \gtrsim 10^{11} {\rm g\ cm^{-3}}$. Therefore, a neutrinosphere is formed, i.e. the radius at which neutrinos are not trapped anymore, and therefore start free-streaming outwards. Notice that, technically speaking, the radius of the neutrinosphere is a function of neutrino energy (higher-energy neutrinos decouple at larger radii), although oftentimes in the literature the term neutrinosphere is used to indicate the energy-averaged neutrinosphere, for simplicity. Moreover, different species will decouple at different radii since they experience different interactions with matter. Specifically, for a given energy, the heavy-lepton neutrinos and antineutrinos decouple at smaller radii than electron antineutrinos, which in turn decouple at smaller radii than electron neutrinos. Given the key role that neutrinos play in the explosion, the transition between the completely trapped regime and the free-streaming regime has to be accurately simulated. This requires solving the Boltzmann transport equation, and several numerical approaches can be chosen, as we will briefly mention in the following sections.

The collapse is halted only when the inner core reaches nuclear saturation densities ($\sim 2.5 \times 10^{14}\ {\rm g\ cm}^{-3}$), at which point the nuclear equation of state stiffens due to the strong force becoming repulsive. The collapse of the inner core stops, and a pressure wave starts propagating outwards. Once this pressure wave reaches the sonic point, it steepens into a shock wave near the edge of the homologous core, and this is what is called the "bounce".

\subsection{The bounce phase}
Once the shock wave is launched, the inner core becomes causally disconnected from the outer layers of the star, and its electron fraction remains practically frozen at a value of $\sim 0.3$. 

The shock propagates outwards and photodissociates the infalling Fe-core material into free nucleons and alpha particles. This quickly drains the kinetic energy of the shock and, after a brief period of positive post-shock velocity \citep{Janka2012_review_CCSNe}, the shock stagnates, and moves outwards in radius simply due to the huge accretion rate that settles matter onto the newly-born Proto-Neutron Star (PNS), thus increasing the post-shock pressure in the form of neutrino heating. The shock reaches a maximum stalling radius of $\sim 150-200$ km after 100--200 ms from the bounce. Now, it will either slowly recede back following the contraction of the PNS, producing a failed SN, or it will be re-energized by some heating mechanism, producing a successful explosion.

\subsection{Neutrino delayed heating}
Once the shock has reached its maximum stalling radius of $\sim 150-200$ km, matter in the post-shock region is emitting neutrinos as a cooling mechanism but is also absorbing neutrinos escaping the PNS. The net effect of these two competing processes is that matter is cooling in the region closer to the PNS, due to neutrino emission being stronger. However, in the region closer to the shock, matter is being heated, since neutrino absorption is stronger than emission. This is the so-called gain region. Therefore, the shock loses energy due to photodissociation of the infalling material, and it gains energy due to neutrino heating. However, all modern spherically symmetric simulations \citep{OConnor2018_comparison} have shown that, without any multidimensional effects, the stalling shock slowly recedes quasi-statically. Eventually, a large fraction of the star will fall back on the central compact object, leaving behind a black hole\footnote{Exceptions to this are extremely low massive stars with steep density profiles\citep{Melson2015_9.6_expl}, or some exotic physics, such as a phase transition to quark matter at high densities \citep{Fischer2021_QCD_massive_SN}}. 

After the neutrino-delayed heating mechanism was established, several studies investigated whether an increased amount of neutrinos emitted from the PNS (mostly due to convective motions below the neutrinosphere) could increase the heating in the gain region enough to trigger an explosion \citep{Epstein1979_lep_conv,Bruenn1979_convective_overturn,WilsonMayle1988}. However, this was later found to be incorrect \citep{Bruenn1996_doubly_diffusive,Bruenn2004_n_fingers_analysis}, although PNS convection is still important in several aspects of CCSNe \citep{Mezzacappa1998_PNS_convection_nu_transport,Roberts2012_PNS_cooling_convection,Nagakura2020_PNS_convection,Pascal2022_PNS_cooling,Akaho2023_PNS_convection_3DGR}. 

\subsection{Multi-dimensional effects}
\label{sec:multi-D_effects}
It is now well-established that, to get a successful explosion, neutrino heating must be complemented by another source of energy that is inherently multi-dimensional. Rotation and magneto-hydrodynamic instabilities have been shown to aid the explosion, although only in a small fraction of CCSNe, due to the large angular momentum and magnetic fields required \citep{Wilson2005_magnetic_rotating_SN,Mosta2014_3D_MRSN,Obergaulinger2020_MRSN_explosion,Bugli2021_MHD_3D_SN}. Another mechanism that has been proposed is the "jittering jets" mechanism \citep{Soker2010_original_jj}, which has however not been observed so far in self-consistent simulations. For most CCSNe, the processes that can aid the explosion are mainly the Standing Accretion Shock Instability (SASI) \cite{Blondin2003,Foglizzo2006_SASI,Marek2009_SASI_diag_ene}, and neutrino-driven turbulent convection \cite{Murphy2013_turb_in_CCSNe,Couch2015_turbulence,Lentz2015_3D,Radice2016,Radice2018_turbulence}.

SASI is a large-scale instability (characterized by $l=1,2$) operating as an advective-acoustic cycle that can amplify shock expansion. Neutrino-driven turbulent convection refers instead to the convective instabilities that originate as a consequence of the negative entropy gradient in the gain region. These two mechanisms are sometimes hard to distinguish numerically (especially in 2D) since they both lead to similar effects on large scales \citep{Muller2020_SNReview}, but have nonetheless very distinct features, the main one being that SASI is a quasi-periodic phenomenon, leading to quasi-periodic oscillations of the shock. 

Since these multi-dimensional phenomena have such an important role in the explosion, several groups have studied how asphericities in the pre-SN star can affect either SASI or neutrino-driven turbulent convection (or both). Generally, they found that density perturbations caused by turbulent motions in the latest stages (i.e. a few minutes) before core-collapse can be accreted through the shock and act as seeds for turbulent motions in the gain region, and therefore facilitate the explosion\citep{Nagakura2013_critical_fluct,Couch2015_3D_final_stages,Muller2016_Oxburning,Nagakura2019_semi_an_prog_asym,Fields2021_3D_burning_precollapse}.

\subsection{Microphysics: the nuclear EOS and neutrino physics}
The bounce and post-bounce phases of the supernova, as well as the deleptonization history during the collapse phase, can dramatically change the following dynamics. Therefore, uncertainties in the nuclear equation of state and neutrino interactions can have a significant impact on the outcomes of a supernova.

As it turns out, the Equation of State (EOS) for nuclear matter is not yet well constrained (e.g. \cite{Hebeler2013_constraints_EOS,Klahn2006_HIC_constraints_EOS,Zhang2018_constraints_on_EOS,Radice2018_contraints_on_EOS_from_GW170817,Burgio2021_constraints_on_EOS,PREX2021_nskin_208Pb,CREX2022_nskin_40Ca}) since the very large densities reached in the PNS are extremely hard to probe. Therefore, several different EOSs satisfy current experimental, observational, and theoretical constraints, despite predicting very different outcomes when it comes to supernova explosions. Given the huge impact that the EOS has on CCSNe (and any other astrophysical phenomena involving compact objects), countless studies have been carried out in the last few decades \citep{Lattimer1991_LS,OConnor2011_explodability,Hempel2012_SN_simulations,Janka2012_review_CCSNe,Steiner2013_SFHo,Couch2013_EOS_dependance,Suwa2013_EOS_dependance,Fischer2014_Sym_Ene_in_SN,Char2015_HyperonEOS_in_SN,Olson2016_NDL_EOS,Furuwawa2017_VM_vs_FYSS_EOSs,Richers2017_EOS_effect_on_GW_SN,Nagakura2018_2D_full_transpo_2EOS,Morozova2018_GW_SN_EOS,Burrows2018_Physical_dependencies_SN,Schneider2019,Harada2020_2D_SN_EOS_effect,Yasin2020_EOS_effects_1DFLASH,Boccioli2022_EOS_effect}. 

In general, soft EOSs tend to facilitate the explosion, whereas stiff EOSs tend to disfavor the explosion. Qualitatively, a soft EOS is characterized by a slow increase of pressure as density increases, whereas for a stiff EOS the pressure increases more rapidly with density. Therefore, for a soft EOS, the PNS contracts to smaller radii since the pressure response to the gravitational pull is small. Therefore, it can reach higher temperatures, which produce more energetic neutrinos, increasing the neutrino heating in the gain region. However, due to the highly non-linear nature of CCSNe, as well as the complicated dependence of neutrino opacities on the EOS, this simple picture is not necessarily realized in simulations. In other words, stiffness is not necessarily correlated with stronger shock revivals and explosions. Nonetheless, there have been some recent papers \citep{Schneider2019,Yasin2020_EOS_effects_1DFLASH} that were able to point out clear correlations between some specific property of the EOS (specifically the effective nucleon mass) and the strength of the explosion. However, this correlation is only valid for EOSs calculated using Skyrme interactions theories. Since the definition (and calculation) of the effective nucleon mass depends on the theoretical framework used to perform the calculation (e.g. Skyrme interaction theories versus relativistic mean-field theories), this correlation breaks down when considering a broader range of EOSs. Interestingly, another correlation can still be observed: the central entropy in the PNS right after bounce correlates with the strength of the subsequent explosion \citep{Boccioli2022_EOS_effect}. No other thermodynamic quantity or neutrino property (i.e. luminosity, energy, neutrinosphere properties, etc...) shows any correlation whatsoever.

\begin{figure}[H]
\begin{subfigure}{0.49\textwidth}
\includegraphics[width=\columnwidth]{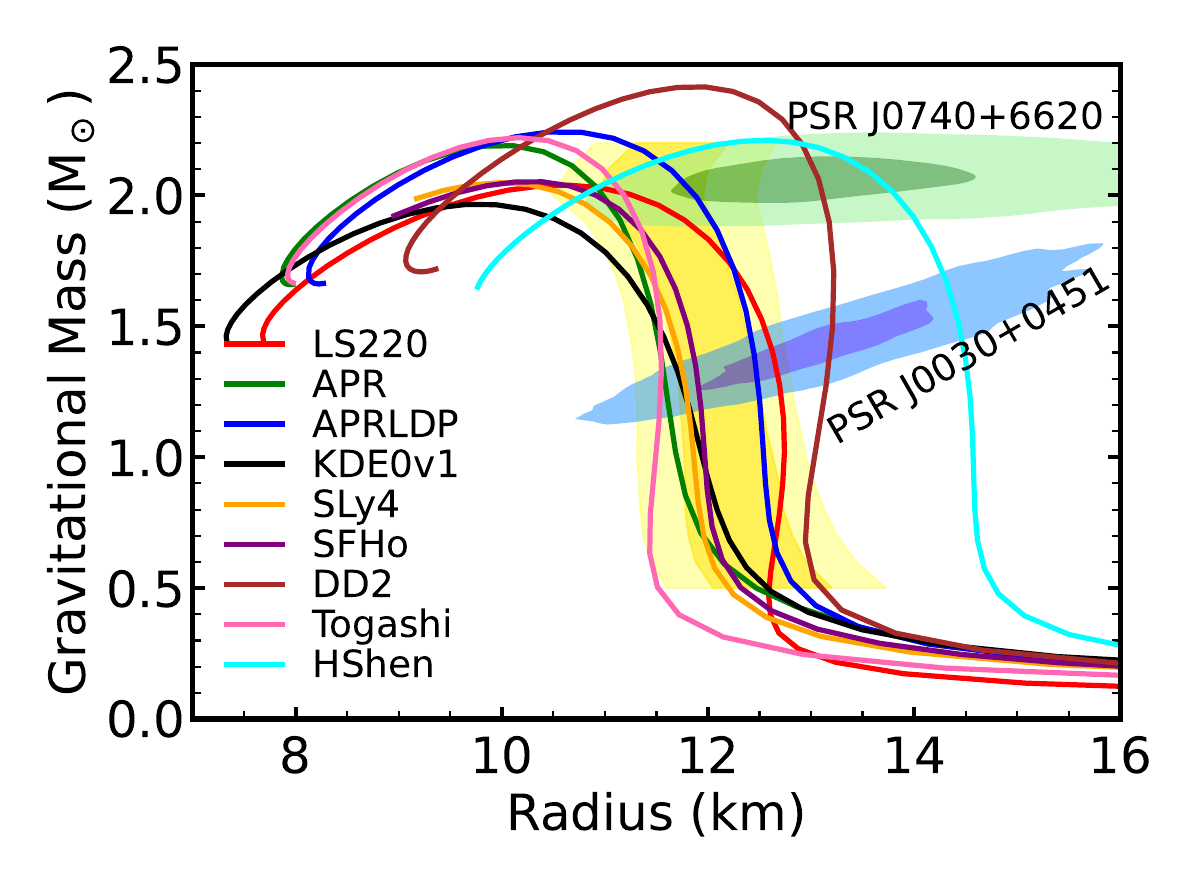}
\end{subfigure}
\begin{subfigure}{0.49\textwidth}
\includegraphics[width=\columnwidth]{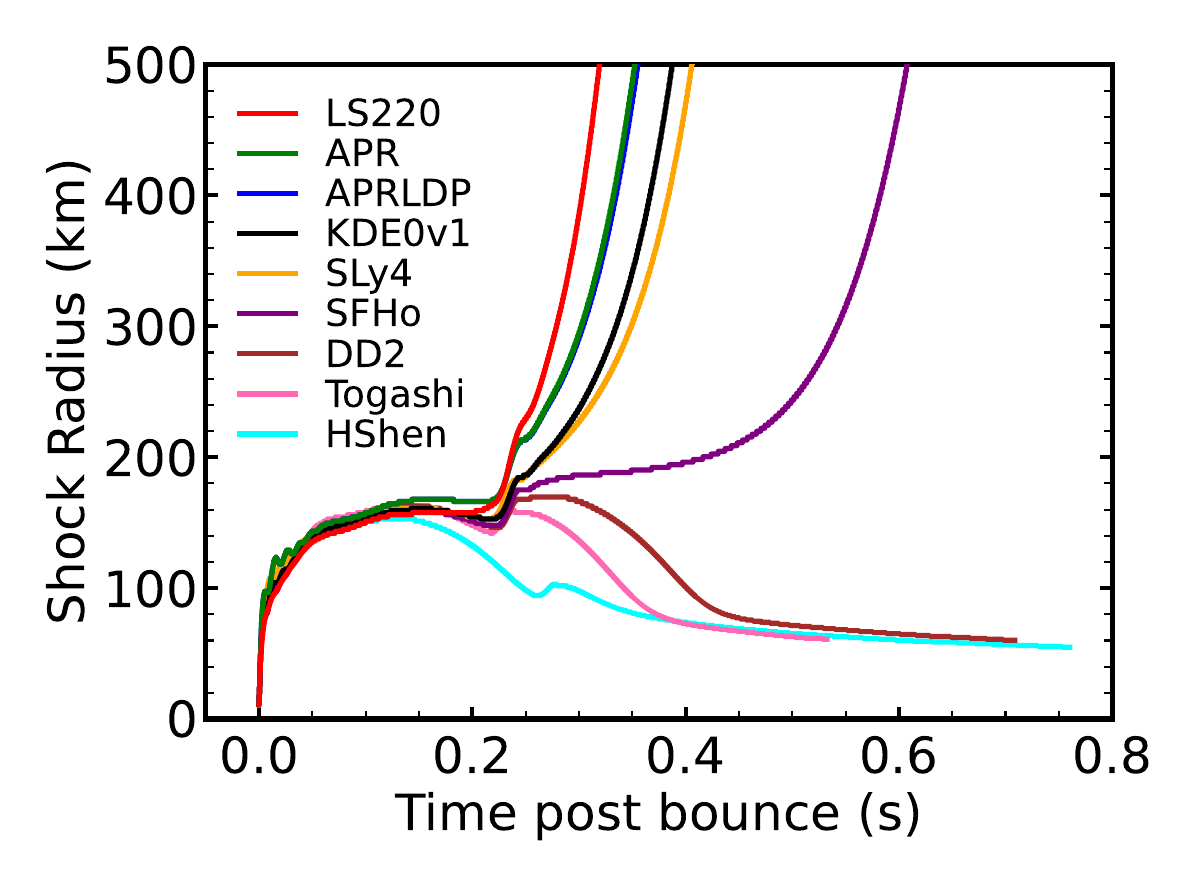}
\end{subfigure}
\caption{The left panel shows Gravitational mass vs radius for cold neutron stars for different EOSs \citep{Lattimer1991_LS,APR_1998,Chabanat1997_SLy4,Hempel2010_HS_RMF,Typel2010_DD2,Shen2011_HShen,Steiner2013_SFHo,Schneider2017_SROEOS,Togashi2017_EOS_real_nuc_pot,Schneider2019_APR_EOS}. The yellow-shaded region shows the constraints from \cite{Nattila2016_EOS_constraints}, whereas the green and blue contours represent the mass and radius of the millisecond pulsars \texttt{PSR J0030+0451} \citep{Miller2019_NICER_measurement} and \texttt{PSR J0740+6620} \citep{Miller2021_NICER_measurement}. The right panel shows the shock evolution of a 20 M$_\odot$ progenitor from \cite{Sukhbold2016_explodability} for the EOSs on the left panel. Simulations were run with the code described in \cite{Boccioli2021_STIR_GR}. See \cite{Boccioli2022_EOS_effect} for more details. \label{fig:EOS_rshk_contraints}}
\end{figure}

To illustrate the wide range of possible outcomes that can be obtained by varying the EOS, we show in \figurename~\ref{fig:EOS_rshk_contraints} the evolution of the shock radius for a 20 M$_\odot$ progenitor for several different EOSs, alongside the mass-radius relations of cold neutron stars allowed by those EOSs. Some of them satisfy all of the current constraints, whereas others have been recently ruled out, but have been widely used in simulations up to the very recent past. It is clear that, especially given the bifurcation nature of the CCSN problem, changes in the EOS can give rise to completely different supernova dynamics and, in certain cases, outcomes.

Another significant source of uncertainty is neutrino physics. Specifically, neutrino transport and neutrino opacities can both largely affect the pre and post-bounce dynamics. Significant progress has recently been made on neutrino opacities \citep{Bruenn1985,Hannestad1998_NN_Brem,Horowitz2002,Buras2003_numutau_pair,BRT2006,Bollig2017_muon_creation,Horowitz2017_virial} as well as neutrino transport \citep{Bruenn1985,Mezzacappa1999_nutransport,Liebendorfer2004_GR_nutransport,Muller2010_coconut_vertex_code,Shibata2011,Cardall2013_GR_M1,OConnor2015,Kuroda2016_GR_nu_transport_code,Nagakura2017_boltz_transport_3D,Richers2017_nu_transport_comparison}. More recent, and significantly harder to interpret in the context of CCSNe, is the role of collective neutrino oscillations \citep{Duan2010,Mirizzi2016,Tamborra2021_flavor_evolution_review}.

Finally, it is worth mentioning the important role that General Relativity plays in CCSNe. Due to the high densities achieved in the PNS, general relativistic effects can be quite significant. Until recently, the vast majority of simulations would solve the Newtonian equations for radiation hydrodynamics, with multipole corrections to the Newtonian potential \textit{à la} \citet{Marek2006}, and only a few multi-dimensional simulations would solve the full GR equations \citep{Muller2012_2D_GR,Akaho2023_PNS_convection_3DGR}. However, current code capabilities and improvements of physical prescriptions implemented, have reached a high enough accuracy that general relativistic effects must be fully accounted for. Some parametric 1D simulations \citep{Boccioli2021_STIR_GR} have already shown that GR might have non-linear effects not reproducible by approximate methods, and therefore full GR simulations will become more and more relevant in the near future.

Due to the complexity of all of these subjects, we point the reader to dedicated reviews \citep{Mirizzi2016,Duan2010,Fischer2017_Review_EOS_nu,Mezzacappa2020_nu_transport_review,Tamborra2021_flavor_evolution_review,Raduta2021_EOS_review,Furusawa2023_EOS_review} and references therein.

\section{CCSNe in spherical symmetry}
Given the prohibitive computational challenge of running hundreds of high-fidelity CCSNe simulations, several studies employ 1D simulations, where the explosion is triggered using a parametric model (although limited 2D studies with up to $\sim 100$ simulations have recently become feasible \citep{Wang2022_prog_study_ram_pressure,Vartanyan2023_nu_100_2D}). There is a plethora of parametric simulations developed over the last thirty years \citep{Blinnikov1993_bomb,woosley:95, Ugliano2012, Perego2015_PUSH1, Ertl2016_explodability, Mabanta2018_MLT_turb, Couch2020_STIR, Boccioli2021_STIR_GR, Fryer2022_nu_conv_remnant_masses,Sasaki2024_STIR_diffusion}, as well as semi-analytic models \citep{Janka2001_conditions_shk_revival,Fryer2012_remnant_popsynth,Pejcha2012_antesonic_condition,Muller2016_prog_connection,Summa2016_prog_dependence_vertex}, aimed at exploring different aspects of supernovae.

In particular, there is a category of recently developed 1D simulations that incorporate neutrino-driven convection through a parametric model based on Reynolds decomposition of the Euler equations and calibrated using multi-dimensional simulations. Such models \citep{Mabanta2018_MLT_turb,Couch2020_STIR,Boccioli2021_STIR_GR,Sasaki2024_STIR_diffusion} are oftentimes referred to as 1D+ models, due to their spherically symmetric nature supplemented by a parametric model for convection, an inherently multi-dimensional phenomenon. As we will show later in this section, some of these models tend to be quite different from previous 1D models, where the explosion is triggered using other methods. They also seem to be in excellent agreement with some 3D simulations across different progenitors, although more detailed studies need to be carried out.

Many of these studies were designed to answer an important question: what causes one star to explode and another one to fail? Two related problems, addressed by some of the same studies, are: (i) whether one can identify a condition that can differentiate between successful shock revival and failed explosion; (ii) whether it is possible to predict the explodability (and explosion properties) based on the structure of the pre-SN progenitor.

\subsection{The critical luminosity condition}
The first attempts at describing an explosion condition date back to the work of \citet{Burrows1993_Theory_SN_expl}. There, the supernova was framed as a bifurcation problem. The only two drastically different solutions are a successful shock revival or a failed explosion. No intermediate solution is possible, and therefore one can imagine that there exists a critical condition that, if met, would cause an explosion. 

As summarized in Section \ref{sec:SN_engine}, the stalling phase of the shock can be described as a balance between neutrino heating and energy lost by the shock due to the photodissociation of the infalling material. Therefore, the two key quantities that describe the shock evolution are the mass accretion rate $\dot{M}$ and the net energy deposited by neutrinos in the gain region in the unit time $\dot{Q}$ (or, equivalently, the neutrino luminosity $L_\nu$). The idea delineated for the first time by \citet{Burrows1993_Theory_SN_expl} is that there is a critical curve in the $\dot{M}$-$\dot{Q}$ plane that, if crossed during the post-bounce phase, would lead to a successful explosion.

Revamped using different formalisms \citep{Thompson2000_tau_crit,Janka2001_conditions_shk_revival,Buras2006_2D_diag_ene,Fernandez2012_timescales,Murphy2017_int_cond}, the concept of a critical condition has been quite successful, and it was confirmed by several multi-dimensional simulations \citep{Murphy2008_crit_lum_2D,Hanke2012_SASI_crit_lum,Fernandez2015_3D_SASI}. However, the existence of a critical condition was usually only qualitatively shown, and when quantitative estimates were given, they were usually only order of magnitude accurate. Recently, the more quantitative Force Explosion Condition (FEC) was derived \citep{Gogilashvili2022_FEC} with the assumption of spherical symmetry, and extensions to multi-dimensions are currently underway \citep{Gogilashvili2023_FEC+}. 

\subsection{The explodability problem} \label{sec:explodability}
Historically, the traditional idea about explodability was that less massive stars $M \lesssim 20$ M$_\odot$ successfully explode and form neutron stars. More massive stars $M \gtrsim 20$ M$_\odot$ lead to failed supernovae and form black holes. This picture has however been challenged by recent studies \citep{Ertl2016_explodability,Sukhbold2016_explodability,Muller2016_prog_connection,Couch2020_STIR,Boccioli2023_explodability,Wang2022_prog_study_ram_pressure} that showed the existence of "islands of explodability" throughout the mass range 9-120 M$_\odot$. However, there is a fundamental disagreement among these studies regarding where these islands are located. The one common trait highlighted in all those papers is the importance of the Si/Si-O interface in determining the outcome of the explosion. 

The first study that provided a quantitative criterion to predict the explodability of massive stars solely based on the pre-SN progenitor, was the one by \citet{Ertl2016_explodability}. Their criterion was based on two quantities: (i) $M_4$, the mass coordinate at which the specific entropy equals 4 (i.e. a proxy for the Si/O interface) and (ii) $\mu_4$ the density gradient at that mass coordinate. That criterion was able to correctly predict the outcome of their 1D simulations. However, as mentioned in Section \ref{sec:SN_engine}, 1D simulations do not self-consistently explode. Therefore, to achieve an explosion, \citet{Ertl2016_explodability} artificially increased the number of neutrinos coming from the PNS following the parametric model of \citet{Ugliano2012}, calibrated on observations of SN1987A \citep{Sonneborn1987a,Bionta1987_nu_IMB,Hirata1987_nu_Kamiokande}. Therefore, their criterion correctly predicted the outcome of those artificially exploded simulations, and it was not clear if that would translate to high-fidelity, self-consistent explosions. From their simulations, they found that progenitors with masses $22-25$ M$_\odot$ and $28-80$ M$_\odot$ lead to failed SN and form black holes.

Later studies \citep{Couch2020_STIR,Wang2022_prog_study_ram_pressure,Tsang2022_ML_explodability,Boccioli2023_explodability} also found islands of explodability, but at very different masses. Two of those studies \citep{Wang2022_prog_study_ram_pressure,Boccioli2023_explodability}, independently derived a criterion to predict the explosion based on properties of the Si/Si-O interface, with a similar approach to \citet{Ertl2016_explodability}. However, only one parameter was used to formulate the criterion. This parameter, in both cases, could be expressed as $\delta \rho^2_* /\rho^2_*$, where $\delta \rho_*$ is the magnitude of the density drop occurring at the Si/Si-O interface, and $\rho_*$ is the density at which the jump occurs. \citet{Boccioli2023_explodability} found explosions when $\delta \rho^2_* /\rho^2_* > 0.08$, whereas \citet{Wang2022_prog_study_ram_pressure} found explosions when $\delta \rho^2_* /\rho^2_* > 0.078$, which is a discrepancy smaller than 3 \%. 

\begin{figure}[H]
\begin{subfigure}{0.49\textwidth}
\includegraphics[width=\columnwidth]{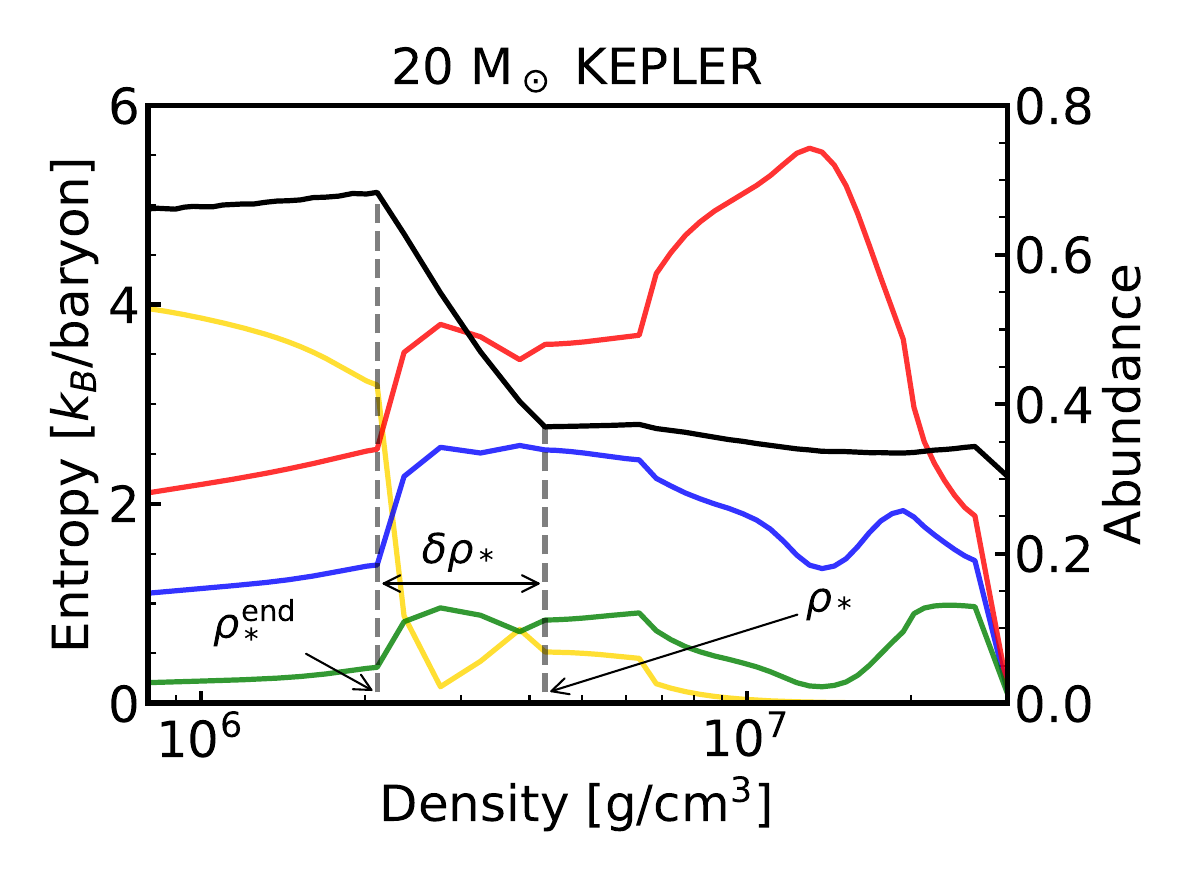}
\end{subfigure}
\begin{subfigure}{0.49\textwidth}
\includegraphics[width=\columnwidth]{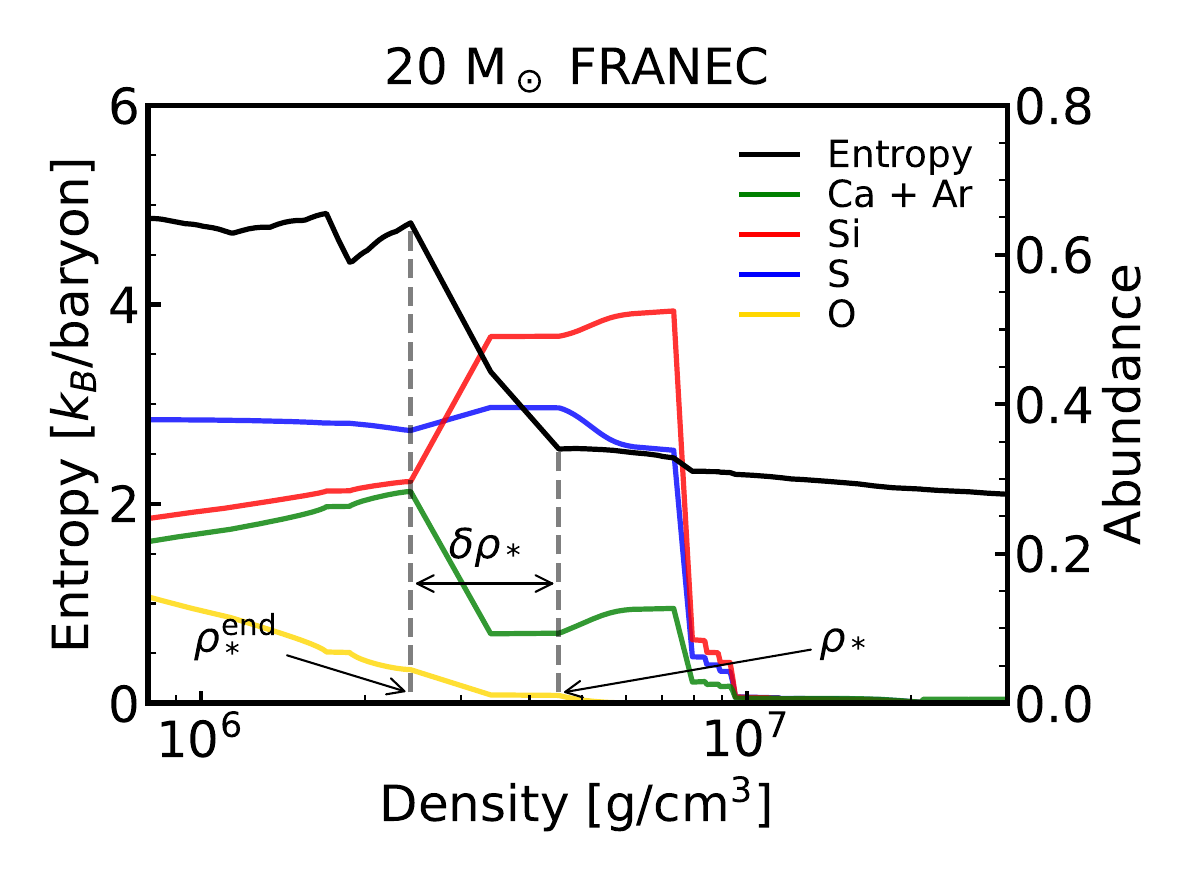}
\end{subfigure}
\caption{Pre-supernova profiles of a 20 M$_\odot$ KEPLER progenitor \citep{Sukhbold2016_explodability} (left) and a 20 M$_\odot$ FRANEC progenitor \citep{LC18} (right). Notice that, despite being quite different, the entropy as a function of density is qualitatively very similar, in particular at the Si/Si-O interface, marked by the vertical dashed lines. Both of these progenitors lead to successful explosions because the interface is located at intermediate densities of $\sim 4 \times 10^6$ g/cm$^3$, and corresponds to a very large density drop of $\delta \rho_*^2 / \rho_* ^2 > 0.2$. The interface also corresponds to a surge in oxygen, although this occurs while the abundance of Si is still quite large, above 0.25. Hence the name Si/Si-O interface. \label{fig:SiO_interface}}
\end{figure}  

The main difference between those two studies is that \citet{Wang2022_prog_study_ram_pressure} used $\sim 100$ 2D simulations, whereas \citet{Boccioli2023_explodability} used $\sim 300$ 1D+ simulations where the explosion was achieved using STIR, a parametric model for $\nu$-driven convection based on Reynolds decomposition and time-dependent mixing-length theory. The main parameter of that model is $\alpha_{\rm MLT}$, as defined in Eq. \eqref{eq:MLT}. The excellent agreement between those two independent studies is a testimony to how 1D+ models can still be employed in some cases without having to compromise on the physics. Moreover, it shows that the Si/Si-O interface represents an important feature of the SN progenitor that has a huge impact on determining the outcome of the explosion. An example of the Si/Si-O interface is shown in \figurename~\ref{fig:SiO_interface} for a 20 \msun\ progenitor from two different stellar evolution codes. The general idea is that, once the Si/Si-O interface is accreted through the shock, the mass accretion rate suddenly drops due to the drop in pre-shock density. This decreases the ram pressure and the shock experiences a temporary expansion that, if supported by sufficient neutrino heating (oftentimes in the form of neutrino-driven convection), can turn into a runaway explosion.

\begin{figure}[H]

\includegraphics[width=\columnwidth]{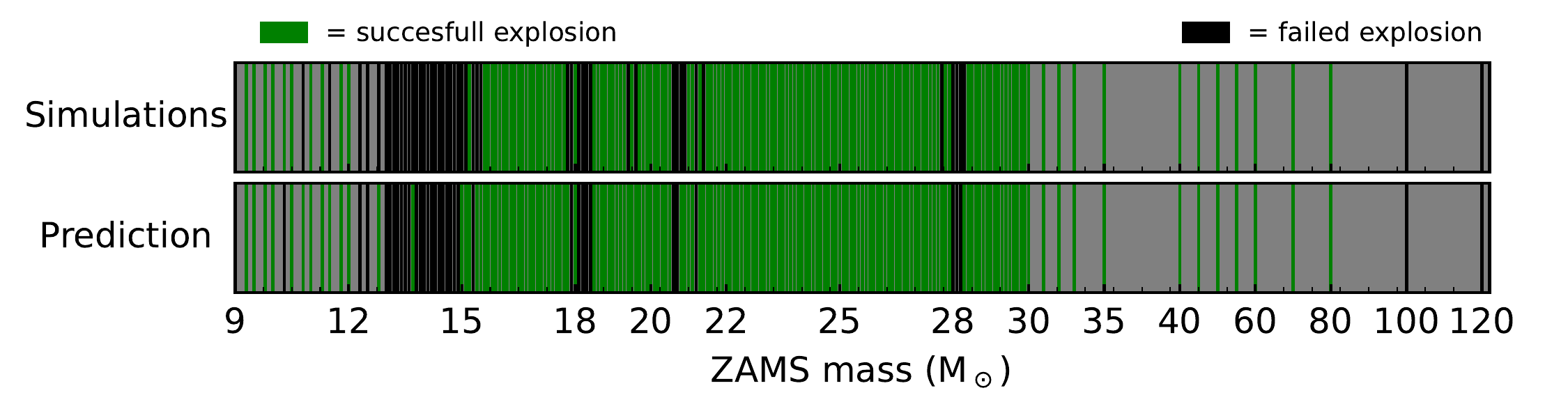}

\caption{Explodability of 200 progenitors from \cite{Sukhbold2016_explodability}. The outcome of the simulation is shown in the upper panel, whereas the prediction according to the criterion of \cite{Boccioli2023_explodability} is shown in the lower panel. All simulations were run using the setup described in \cite{Boccioli2023_explodability}, with $\alpha_{\rm MLT} = 1.51$. \label{fig:explodability_criterion}}
\end{figure}

\begin{figure}[H]

\includegraphics[width=\columnwidth]{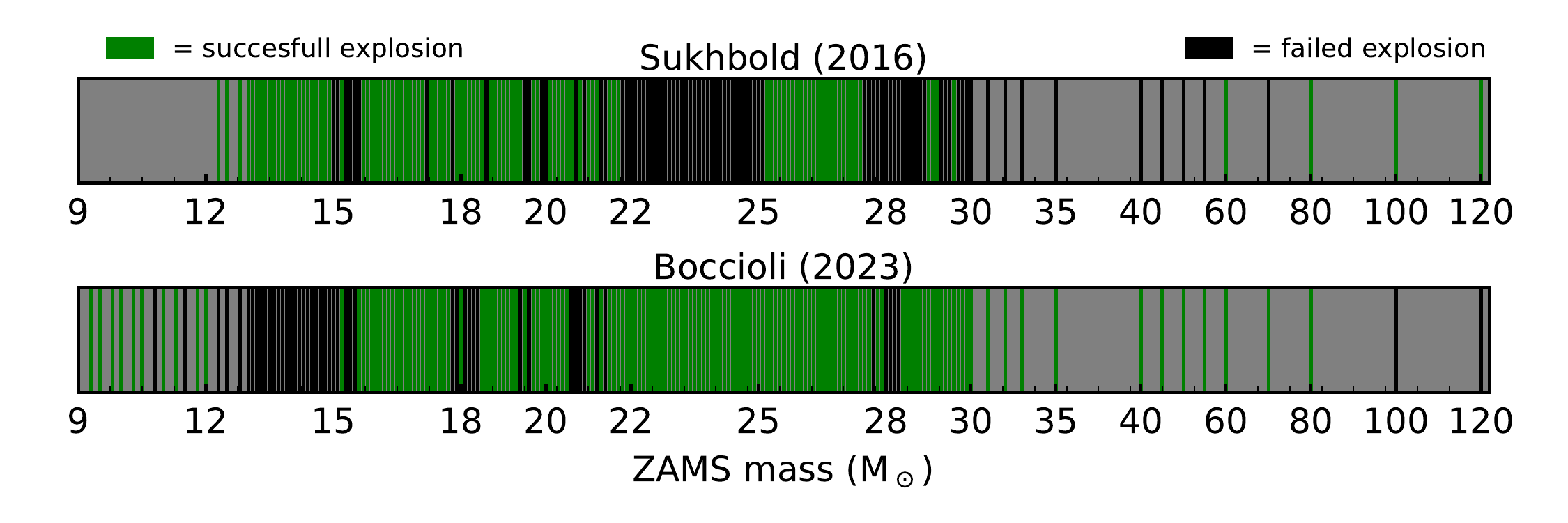}

\caption{Comparison between the explodability found by \citet{Boccioli2023_explodability} and the one found by \citet{Sukhbold2016_explodability} (which is the same as the one found by \citet{Ertl2016_explodability}). The upper panel shows the results from the $\alpha_{\rm MLT} = 1.51$ model of \cite{Boccioli2023_explodability}, whereas the lower panel shows the results from the N20 model of \cite{Sukhbold2016_explodability}. The black-hole-forming regions are very different, the reason being that the explosion is triggered in different ways. See \citet{Boccioli2023_explodability} for a detailed discussion. \label{fig:explodability_N20_vs_STIR}}
\end{figure}

The explodability prediction of the criterion derived by \citet{Boccioli2023_explodability} (but in the case of \citep{Wang2022_prog_study_ram_pressure} it would be quite similar) is shown in \figurename~\ref{fig:explodability_criterion} for $200$ 1D+ simulations from \citep{Sukhbold2016_explodability}. The criterion correctly predicts the outcome of the simulations in more than $90 \%$ of progenitors and recovers the islands of explodability extremely well. One can see that the progenitors forming black holes are mostly the low-mass ones with $12 < M < 15$ M$_\odot$, as well as a few around 18, 28, and $\gtrsim 100$ M$_\odot$. The progenitors used are the same as the ones employed by \cite{Ertl2016_explodability} and \cite{Sukhbold2016_explodability}. However, the explodability derived is quite different, as shown in \figurename~\ref{fig:explodability_N20_vs_STIR}. The black-hole-forming regions are completely different, which highlights how large the uncertainties in the explosion mechanism are, and it shows that multi-dimensional simulations are needed to determine whether parametric 1D and 1D+ simulations correctly model the explosion. As mentioned in the previous sections, large sets of multi-dimensional simulations have now become feasible, and the explodability found by \citet{Boccioli2023_explodability} and \citet{Wang2022_prog_study_ram_pressure} generally agrees with the results of a dozen high-fidelity 3D simulations performed by \citet{Burrows2020_3DFornax}. This shows how 1D+ simulations can now well reproduce the qualitative general trend and evolution of high-fidelity multi-dimensional simulations, and can therefore still prove to be useful tools to explore the parameter space of supernovae.

\subsubsection{Impact of stellar evolution uncertainties on the explodability}
\begin{figure}[H]

\includegraphics[width=\columnwidth]{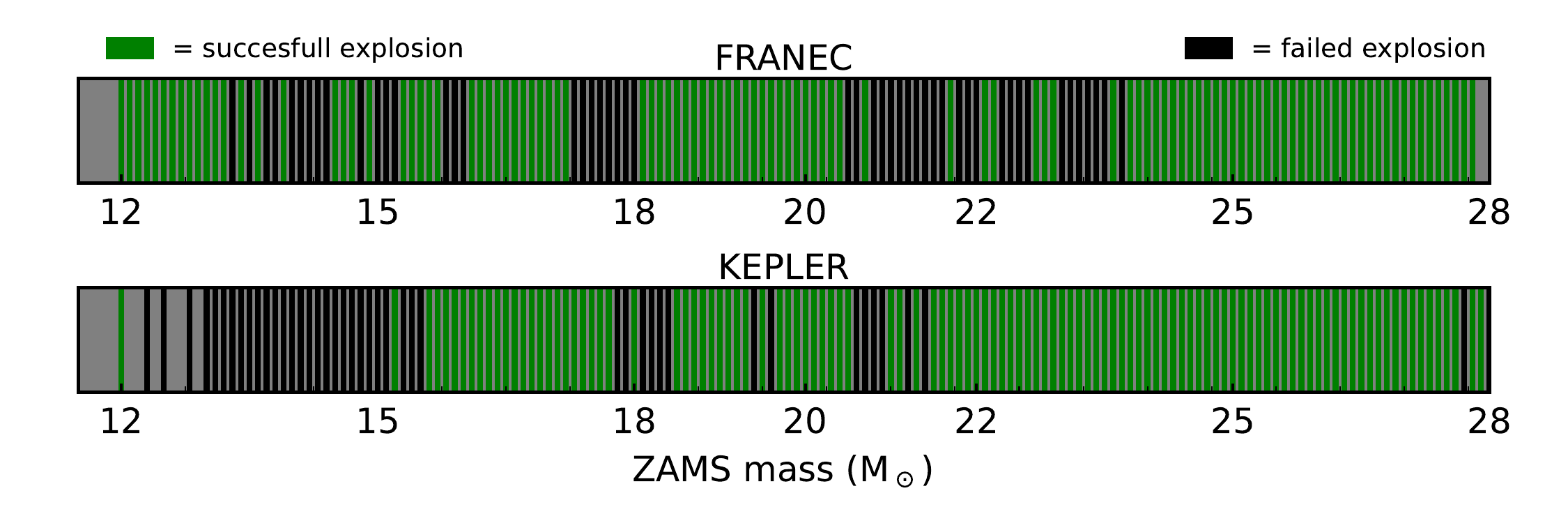}

\caption{Figure adapted from \cite{Boccioli2023_explodability}. The upper panels show the outcome of simulations run using the stellar evolution code FRANEC \citep{Chieffi2020_presupernova_models}, whereas the lower panel shows the outcome of simulations run using the stellar evolution code KEPLER \citep{Sukhbold2016_explodability}. The details of the simulations can be found in \cite{Boccioli2023_explodability}. \label{fig:explodability_stev}}
\end{figure}

It is important to point out that the supernova is only sensitive to the stellar profiles right before collapse. That is, it is not directly related to the Zero-Age Main Sequence (ZAMS) mass and metallicity of the star. Therefore, as seen in Sect. \ref{sec:codes}, differences in stellar evolution calculations will yield different pre-collapse profiles, and therefore different explosion outcomes. An example of this can be seen by comparing the upper and lower panels of \figurename~\ref{fig:explodability_stev}, adapted from \cite{Boccioli2023_explodability}. The upper panel shows the explodability of stars computed using the stellar evolution code FRANEC
; the lower panel shows the explodability of stars computed using the stellar evolution code KEPLER \citep{Sukhbold2016_explodability} (lower panel). 

A detailed comparison between codes is extremely challenging given the differences in algorithms, assumptions, and overall physical processes considered. It is however apparent that there are many uncertainties in the stellar evolution of massive stars (see Sect. \ref{sec:uncertainties}), and that these uncertainties hugely affect the final structure of a star, which in turn changes the explodability. Moreover, multi-dimensional effects during the evolution of massive stars (especially in the last burning stages) are not well understood, as mentioned in Section \ref{sec:multi-D_effects}. Finally, the explodability discussed here assumes single-star evolution, but we know that about two-thirds of massive stars are in binary systems, and that can also significantly change the final structure of the star, and therefore the outcome of the explosion \citep{laplace:21,Vartanyan2021_Binary_stars_SiO_interface}.

\section{Propagation of the shock and explosive nucleosynthesis}

In the case of a successful explosion, the re-energized shock escapes the Fe core and is free to propagate through the mantle of the star, accelerating and pushing away the stellar material above the forming PNS. The result is that part of the envelope is ejected and pollutes the surrounding interstellar medium, while the innermost layers above the Fe core fall back onto the PNS because of its strong gravitational field, accreting on the central compact object. Due to the lack of a full self-consistent theory for the final stages of the hydrostatic evolution of a massive star, the core-collapse, and the post-bounce evolution (see previous sections), the late stages of CCSNe are usually simulated by injecting a certain amount of energy at the outer edge of the Fe core, and then following the propagation of the shock with semi-analytical prescriptions or hydrodynamic simulations. 

The injected energy is often calibrated to obtain a kinetic energy of the ejecta of the order of $10^{51}\rm erg$ (1 foe) and it may be injected in the form of thermal energy (thermal bomb), kinetic energy (kinetic bomb), or by giving a certain velocity to the outer edge of the Fe core (piston). Usually, in a CCSN calculations, the mass-coordinate that separates the ejecta from the fallback is called mass-cut and it is an arbitrary quantity. It can be fixed at the end of the simulation to eject a certain amount of \isotope[56]{Ni} \citep[e.g.][]{LC18}, by using interpolation formulae \citep[as used, e.g., by][]{ritter:18}, or as the result of the hydro calculation \citep[][]{rauscher:02,Fryer2012_remnant_popsynth,sieverding:18,curtis:19}. Regardless of the method used to simulate the supernova explosion, the passage of the shock wave has significant effects on the stellar material above the Fe core.

\subsection{Explosive nucleosynthesis}

\begin{figure}[!t]
\includegraphics[width=\columnwidth]{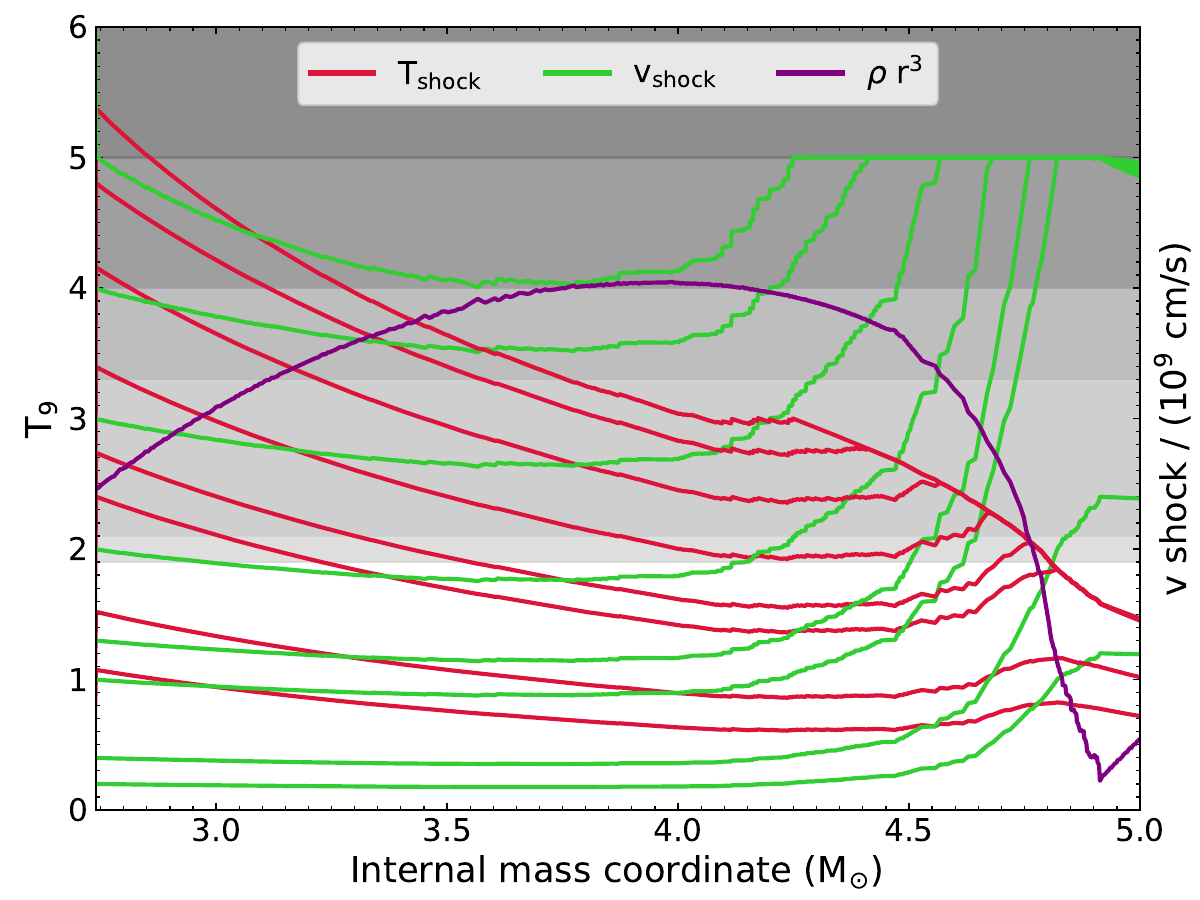}
\caption{Peak shock temperature (red solid lines) and velocity (green solid lines) for different initial conditions in a 20 \msun\ progenitor at solar metallicity \citep{ritter:18}. The purple line represents the logarithm of the quantity $\rho r^3$ in the pre-supernova model in arbitrary units. \label{fig:tpeak}}
\end{figure}

During its propagation through the stellar structure, the shock locally induces compression and heating. The sudden variation of temperature and density in each layer of the star triggers the so-called "explosive nucleosynthesis", i.e. a variation of the chemical composition due to nuclear reactions occurring on a timescale of a few seconds. It is possible to derive the basic properties of the explosive nucleosynthesis without the aid of hydrodynamic simulations, under the assumption that the shock is adiabatic, radiation dominated, and it is propagating in spherical symmetry. The relation between the densities in the regions behind and after the shock wave, in case of strong shock limit \citep{chevalier:89,pignatari:16}, is: 

\begin{equation}
\rho_{\rm shock}=\frac{\gamma+1}{\gamma-1}\rho_{\rm pre};
\end{equation}

with $\gamma$ the adiabatic index. For two typical values of $\gamma$, i.e., 5/3 and 4/3, we obtain $\rho_{\rm shock}=4\rho_{\rm pre}$ and $\rho_{\rm shock}=7\rho_{\rm pre}$, respectively. A similar relation holds for the temperature: 

\begin{equation}
T_{\rm shock}\propto {\rm Ma}^2T_{\rm pre};
\end{equation}

where ${\rm Ma}\gg1$ is the Mach number in the case of strong shock. Hence, the passage of the shock induces a compression that keeps the density almost of the same order of magnitude as in the pre-shock zones, while simultaneously generating a very strong heating of the matter crossed. Furthermore, the relation:


\begin{equation}
    \begin{split}
    T_{\rm shock} &= \left( \frac{3}{2a}\right)^{1/4} [(\gamma+1)\rho_{\rm pre}]^{1/4} v_{\rm shock}^{1/2} \\
    &= 3.8\times10^9 \cdot \left(\frac{\rho_{\rm pre}}{10^7 {\rm g/cm}^3}\right)^{1/4} \cdot \left(\frac{v_{\rm shock}}{2\times10^9 {\rm cm/s}}\right)^{1/2} \ \rm K
    \end{split}
    \label{eq:t}
\end{equation}
shows that the velocity of the fluid hit by the shock is directly related to the shock temperature. In other words, the kinetic energy is converted into internal energy and heats up the matter. Note that $(\gamma+1)^{1/4} \sim 1.25$, since the adiabatic index is always between 4/3 and 5/3. \figurename~\ref{fig:tpeak} shows how different shock velocities lead to different peak temperatures in a semi-analytical simulation. It is worth mentioning that the shock wave accelerates where the quantity $\rho r^3$, i.e., the density of the star multiplied by the radius to the power of three, decreases \citep{Woosley2002_KEPLER_models}. This acceleration occurs when the shock wave encounters the interface between the CO core and the He layer. Hydrodynamic simulations show that \citep{Bethe1990_SN_review,limongi:03a,Herwig2000}, as a consequence of the density contrast between the inner CO-core and the outer He layer, a reverse shock (i.e. a shock propagating inward) is formed at this interface. However, the effect of this reverse shock on the amount of fallback is negligible.

\begin{figure}[!t]
\includegraphics[width=\columnwidth]{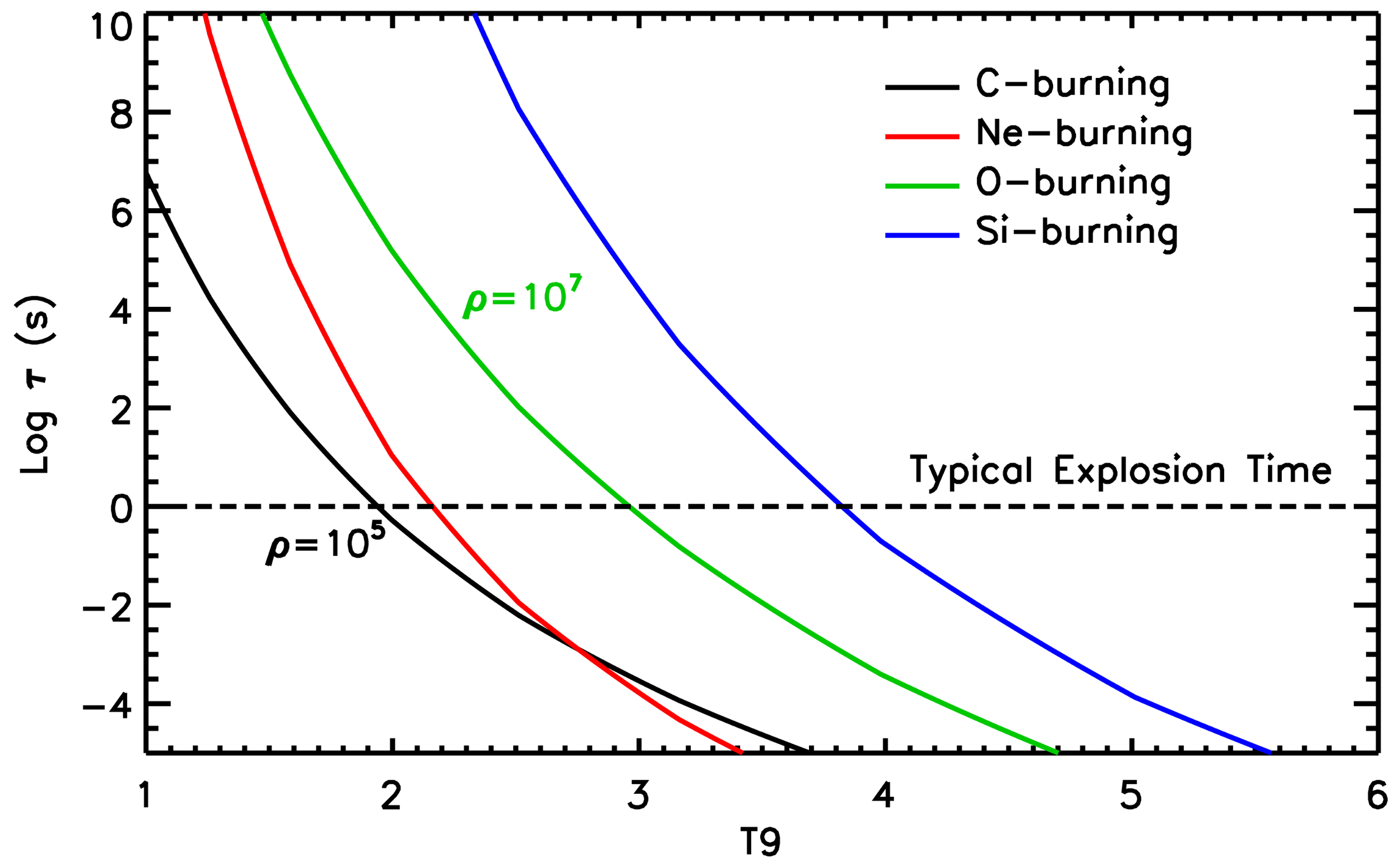}
\caption{Nuclear burning timescale $\tau$ for the main destruction channel of Si (solid blue line), O solid green line), Ne (solid red line), and C (solid black line) as a function of the temperature in units of $10^9$ K ($T_9$). The horizontal dashed line corresponds to $\tau=1$, i.e., the typical explosion timescale. $\rho$ indicates the typical density in the star relative to C-rich zones (black) and to O-rich zones (green) in $\rm g\ cm^{-3}$. The temperature required to burn a significant fraction of each fuel on the explosion timescale are $4\times10^9$ K for \isotope[28]{Si}, $3.3\times10^9$ K for \isotope[16]{O}, $2.1\times10^9$ K for \isotope[20]{Ne}, and $1.9\times10^9$ K for \isotope[12]{C}. \label{fig:tau}}
\end{figure}

The compression and the heating induced by the passage of the shock wave trigger a rapid nucleosynthesis that occurs within seconds, much faster than the characteristic timescale of the pre-supernova nucleosynthesis. In the post-shock zones, pressure, density, and temperature are almost constant in mass and they decrease as matter expands and cools down. In order to determine the typical explosive burning temperatures we require that the nuclear burning timescale $\tau$ for each major stellar fuel (namely H, He, C, Ne, O, and Si, see Sect. \ref{sec:evo}) must be comparable with the explosion timescale. Assuming $\tau\sim1$ s and considering only the principal channel of destruction for each fuel, it is possible to find the minimum temperature required to burn a considerable fraction of a given element, since the explosive timescale depends just on temperature and density. The variation of $\tau$ as a function of $T_9=T/10^9$ K is shown in \figurename~\ref{fig:tau}. From the Figure, we note that the threshold temperatures are $4\times10^9$ K for \isotope[28]{Si}, $3.3\times10^9$ K for \isotope[16]{O}, $2.1\times10^9$ K for \isotope[20]{Ne}, and $1.9\times10^9$ K for \isotope[12]{C}. For temperatures $T_{\rm shock}<1.9\times10^9$ K the timescale for nuclear burning is too long compared to the explosion timescale to trigger any relevant nuclear reaction.

Using these threshold temperatures it is also possible to define the geometrical volumes in which these burning stages occur. Since the shock is radiation dominated, we can write the radius of the shock as a function of the explosion energy and the peak temperature only:

\begin{equation}
R = \left(\frac{3E_{\rm exp}}{4\pi a} \right)^{1/3} T_{\rm peak}^{-4/3}    
\label{eq:radius}
\end{equation}

where R is the radial coordinate, $E_{\rm expl}$ is the explosion energy, and $a$ is the radiation density constant. Equation (\ref{eq:radius}) identifies the spherical volumes within which each explosive burning stage occurs and these volumes depend just on the explosion energy. The specific amount of mass that is involved in the explosive nucleosynthesis within each volume depends on the mass-radius relation in the PSN model, while the relative proportions among the nucleosynthesis products depend both on the $Y_e$ profile and on the pre-supernova abundances. 

\subsection{Explosive burning stages} \label{sec:expburn}

The explosive Si burning occurs in all the zones heated up to $T_9=4$. For temperatures $T>5\times10^9$ K, a full nuclear statistical equilibrium is achieved, i.e., a unique cluster of elements in equilibrium is formed and all the Si is converted into light particles ($p$, $n$, and \A) and later into Fe peak nuclei. Their relative proportions are a function of temperature, density, expansion (cooling) timescale, and local neutron excess. This stage is called "complete Si burning". If the timescale of the expansion is shorter than the timescale required to reach NSE, the chemical composition is dominated by \A\ nuclei (\A-rich freeze-out). Otherwise, the main products are Sc, Ti, Co, Zn, and Ni. In particular, \isotope[56]{Ni} is the most abundant nuclear specie, and it is the main driver of the supernova light curve, through its decay to \isotope[56]{Co} first and to \isotope[56]{Fe} then.

If the peak temperature is $4\times10^9\ {\rm K}<T<5\times10^9\ {\rm K}$, not all the processes are in statistical equilibrium, and two equilibrium clusters of elements are formed. They are defined by $A>46$ and $A<44$, respectively, with $A$ being the atomic mass. Since the matter crossed by the shock at these temperatures is rich in elements with $A\ll44$, the value $A=44$ is a bottleneck for all the reactions that produce elements with higher $A$. As a consequence, Si is only partially exhausted and the remaining Si abundance decreases with temperature. For this reason, this stage is called "incomplete Si burning". In this case, the most produced isotopes are Ca, Ar, and Si. A few processes may overcome the $A=44$ barrier and produce V, Cr, Mn, Fe, and Ni.

In the temperature range $3.3\times10^9\ {\rm K}<T<4\times10^9\ {\rm K}$ only one equilibrium cluster is established at $24<A<44$. Therefore, also in this case $A=44$ constitutes a barrier for the production of heavier elements. This temperature range is usually reached in O-rich zones, and the most produced nuclear species are \isotope[28]{Si}, \isotope[32]{S} \isotope[36]{Ar}, and \isotope[40]{Ca}. 

A further decrease in temperature, down to $1.9\times10^9\ {\rm K}<T<3.3\times10^9\ {\rm K}$, does not allow the formation of any cluster of equilibrium. Therefore, explosive Ne ($T>2.1\times10^9$ K) and C ($T>1.9\times10^9$ K) burning occur at the same pace as during the hydrostatic evolution of the star. The main explosive products are \isotope[24]{Mg}, \isotope[27]{Al}, \isotope[29,30]{Si}, \isotope[31]{P}, and \isotope[37]{Cl} for explosive Ne burning and \isotope[20]{Ne} and \isotope[23]{Na} for explosive C burning.

Below $1.9\times10^9$ K the temperature is not high enough, relative to the explosion timescale, to trigger further major nuclear burning. However, in the case of highly energetic supernovae, some additional nucleosynthetic processes may occur, such as the activation of \isotope[22]{Ne}(\A,$n$)\isotope[25]{Mg} in the He shell, which is crucial for the interpretation of the abundances measured in pre-solar silicon carbide grains \citep{pignatari:18}. 

\begin{figure}[!t]
\includegraphics[width=\columnwidth]{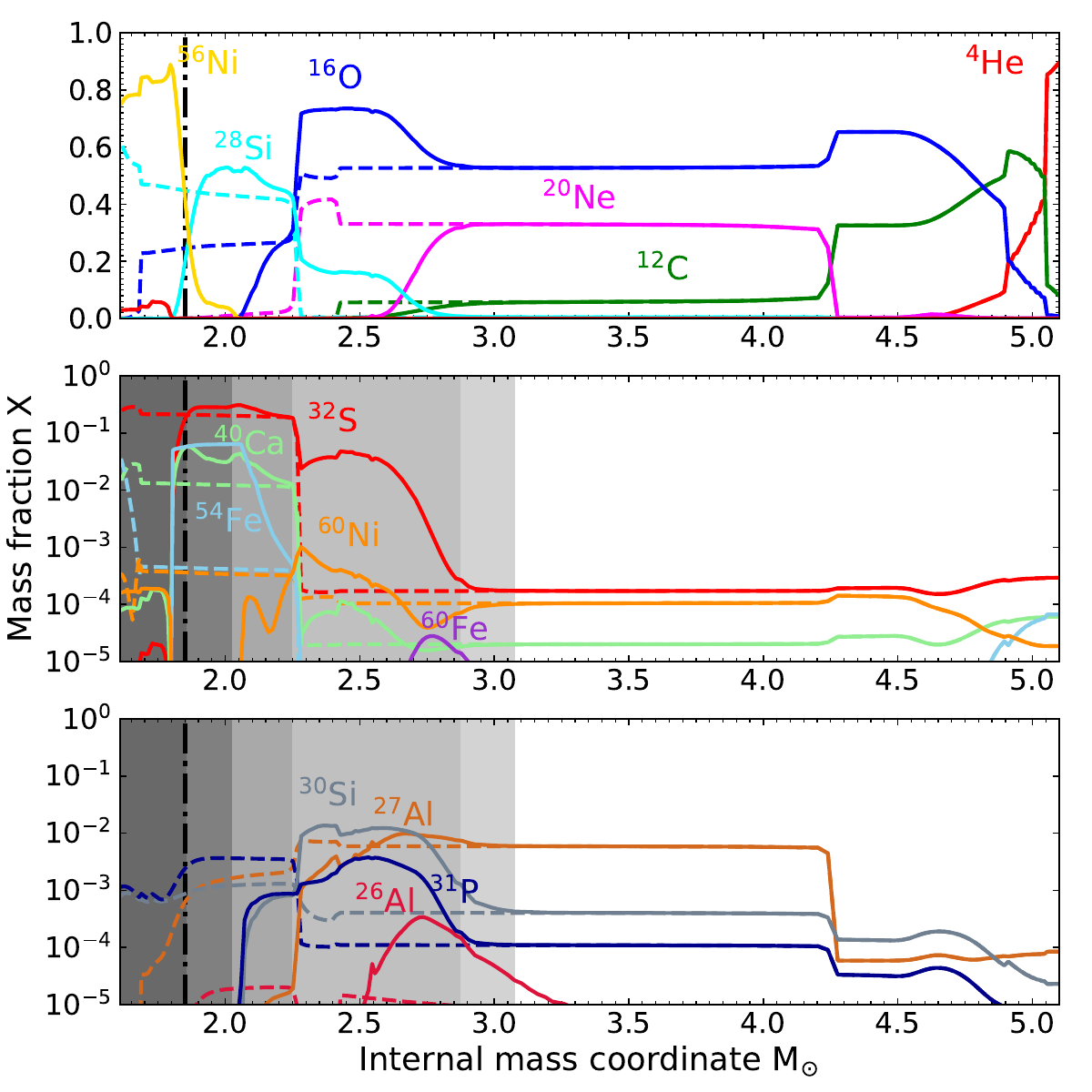}
\caption{Comparison between the pre-supernova (dashed lines) and post-supernova (solid lines) chemical composition in a 20 \msun\ star model exploding with an energy of 2 foe (see text). The upper panel shows the effect of the supernova on the major fuels. The central and the lower panels show instead the effect of the explosive nucleosynthesis on secondary nuclear species. The vertical dot-dashed line represents the location of the mass-cut, chosen to eject a mass of 0.07 \msun\ of \isotope[56]{Ni}. \label{fig:strut}}
\end{figure}

\begin{figure}[!t]
\includegraphics[width=\columnwidth]{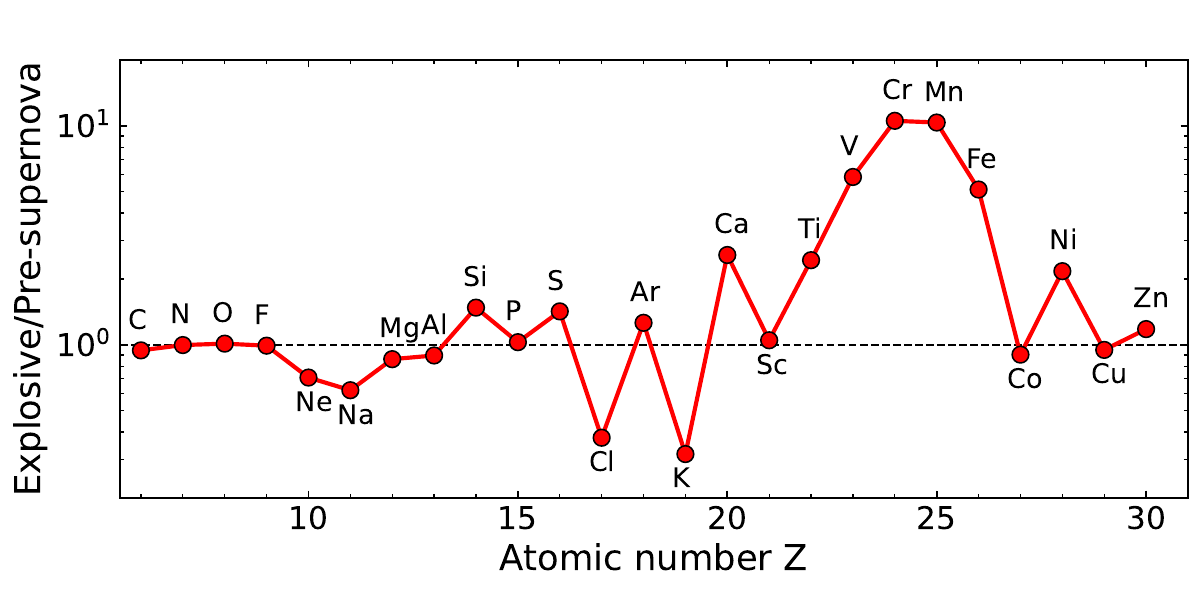}
\caption{Ratio between the explosive and the pre-supernova yields in the model shown in \figurename~\ref{fig:strut}. The explosion mainly produces the elements between Ca and Zn. Elements heavier than Zn are not shown in the figure because they are not significantly affected by explosive nucleosynthesis. \label{fig:yields}}
\end{figure}

\figurename~\ref{fig:strut} shows the explosive nucleosynthesis production within the stellar structure of a selected number of nuclear species for the same FRANEC model shown in \figurename~\ref{fig:psn}. The explosion in this case is simulated in the HYPERION code \citep{LC20} with an explosion energy of 2 foe. The geometrical volumes corresponding to each explosive burning stage (Equation \ref{eq:radius}) are represented as grey shaded areas in the figure and are also reported in \tablename~\ref{tab:radii}. This figure clearly shows that the chemical composition is deeply modified by the explosive nucleosynthesis up to a fraction of the C convective shell. The ratio between the explosive and the pre-supernova yields is instead shown in \figurename~\ref{fig:yields}. We remind the reader that the yield of a certain element is the integral of its abundance over the mass of the star, from the mass-cut up to the surface and including the wind contribution. The mass-cut is such that the ejecta contains 0.07 \msun\ of \isotope[56]{Ni} and corresponds to 1.85 \msun. The explosion mainly produces elements between Ca and Zn, whereas, for all of the other elements, the main contribution comes from the hydrostatic evolution of the star. 

\begin{table}[!t]
\centering
\caption{Explosive burning stages and corresponding volumes in a 20 \msun\ star at solar metallicity exploding with $E_{exp}=2$ foe.}
 \label{tab:radii}               
 \begin{tabular}{rllr}        
 \hline\hline           
 Stage & Temperature & Mass    & Radius \\
           &  (GK)             & (\msun) & (km) \\
 \hline
 complete Si    & $>5$    & 1.86 & 4657 \\
incomplete Si  & $>4$    & 2.03 & 6270 \\
explosive O     & $>3.3$ & 2.25 & 8104 \\
explosive Ne   & $>2.1$ & 2.88 & 14806 \\
explosive C     & $>1.9$ & 3.07 & 16919 \\
\hline          
\end{tabular}    
\end{table}

\subsection{The production of \p-nuclei: the $\gamma$-process in CCSNe}

\begin{figure}[!t]
\includegraphics[width=\columnwidth]{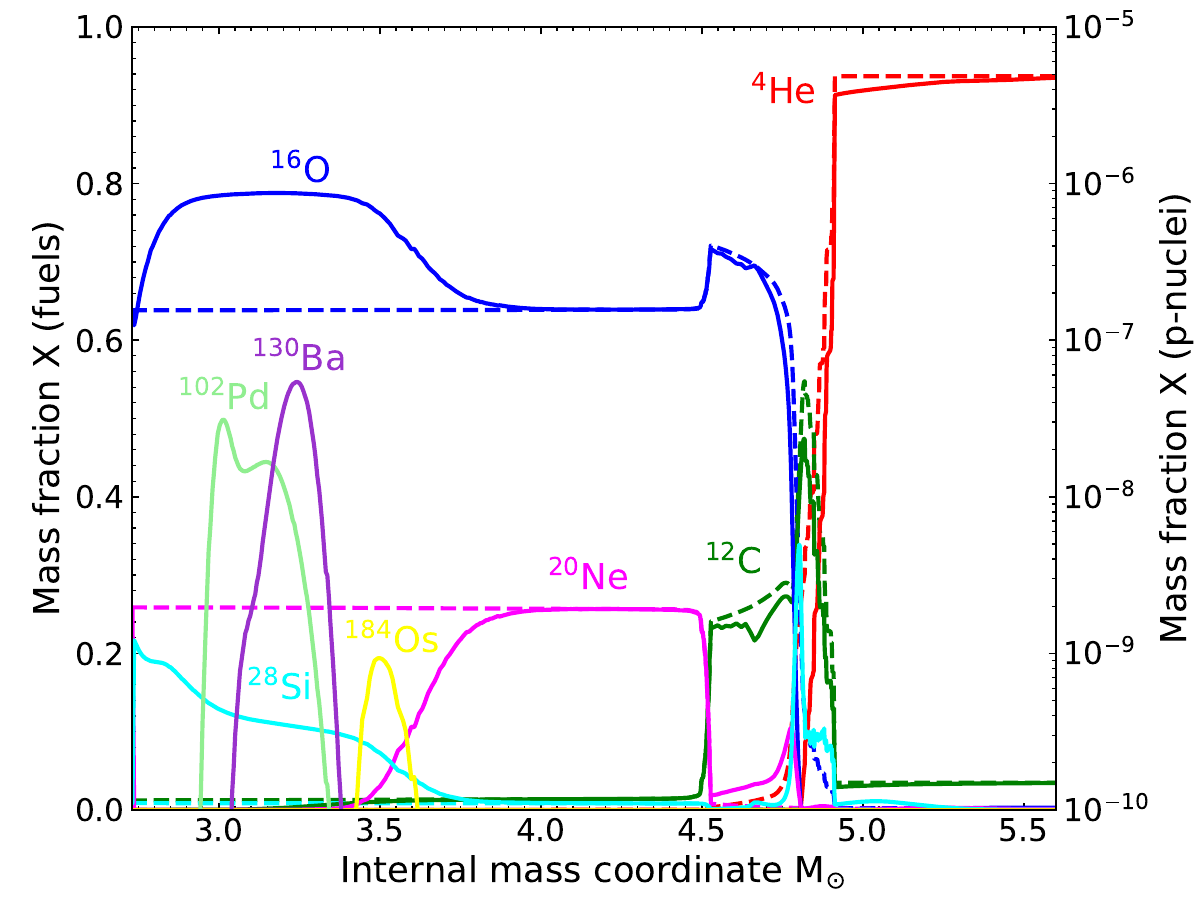}
\caption{Same as \figurename~\ref{fig:strut}, but in the case of the MESA model shown in \figurename~\ref{fig:psn}. The major fuels refer to the left axis, the \p-nuclei abundances refer to the right logarithmic axis. \label{fig:gammap}}
\end{figure}

CCSNe are also one of the possible production sites of 35 rare neutron-deficient isotopes of elements heavier than Fe\footnote{namely: \isotope[74]{Se}, 
    \isotope[78]{Kr}, \isotope[84]{Sr}, \isotope[92,94]{Mo},     \isotope[96,98]{Ru}, 
    \isotope[102]{Pd}, \isotope[106,108]{Cd}, \isotope[112,114,115]{Sn}, 
    \isotope[113]{In}, \isotope[120]{Te}, \isotope[124,126]{Xe},     \isotope[130,132]{Ba}, 
    \isotope[136,138]{Ce}, \isotope[138]{La}, \isotope[144]{Sm},     \isotope[152]{Gd}, 
    \isotope[156,158]{Dy}, \isotope[162,164]{Er}, \isotope[168]{Yb},     \isotope[174]{Hf}, 
    \isotope[180]{Ta}, \isotope[180]{W}, \isotope[184]{Os},     \isotope[190]{Pt},
    and \isotope[196]{Hg}.}
, the so-called \p-nuclei \citep{woosley:78,prantzos:90b,rayet:95,arnould:03,pignatari:16a,roberti:23a}. These peculiar nuclei can not be synthesized by neutron capture processes. Instead, they are produced via chains of photodisintegrations on trans-iron seeds in O/Ne--rich layers of exploding massive stars (\g--process), at temperatures ranging between 2 and 3.5 GK. \figurename~\ref{fig:gammap} shows the abundances of three typical \p-nuclei after the supernova explosion of the MESA model (published in \citep{ritter:18}) already presented in Sect. \ref{sec:evo} and in the upper panel of \figurename~\ref{fig:psn}. Note that the mass region where the \g--process occurs coincides with the explosive Ne burning zone.

\begin{figure}[!t]
\includegraphics[width=\columnwidth]{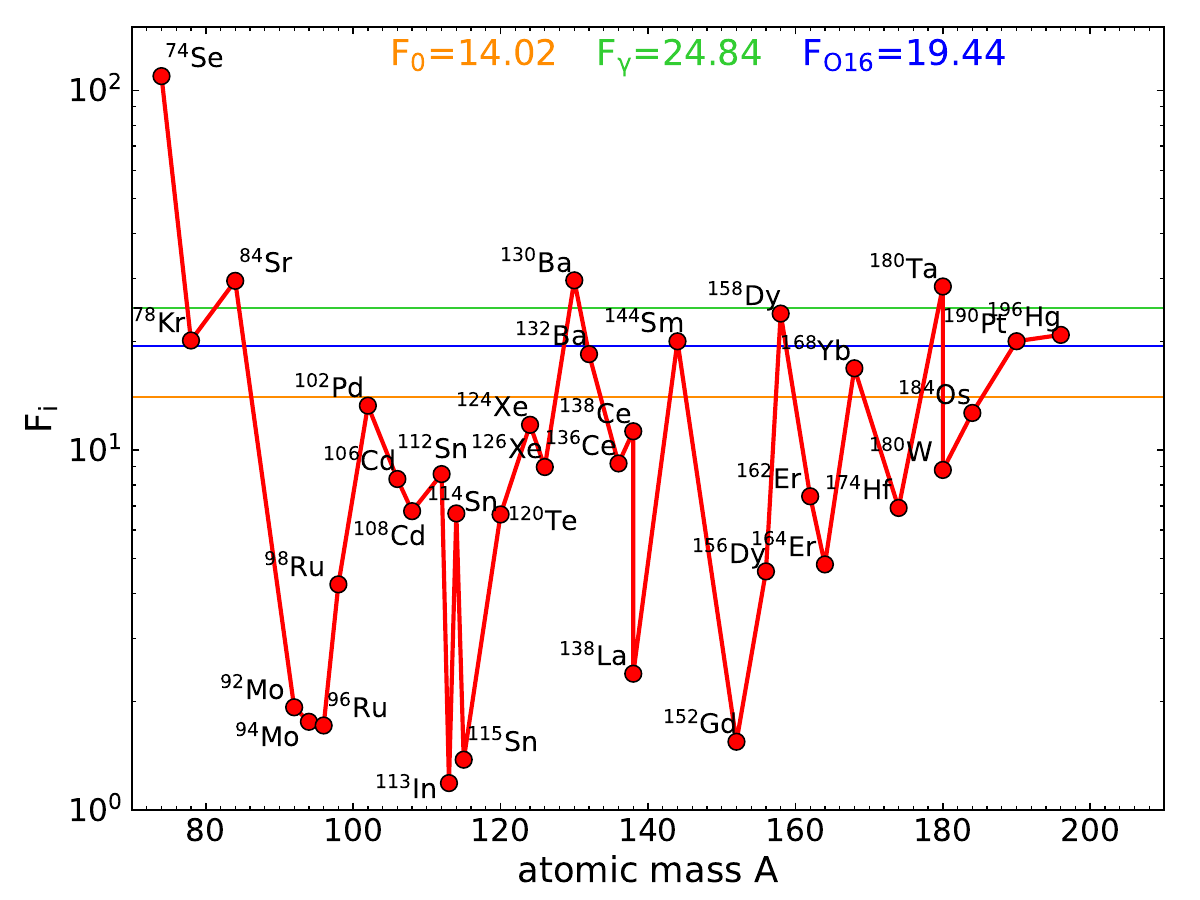}
\caption{Over-production factors of the 35 \p-nuclei. The orange horizontal line represents the average over-production factor \fz, the horizontal green line represents the \g-only factor \fg, and the horizontal blue line represents the oxygen over-production factor, \fo. Most of \p-nuclei are under-produced compared to \isotope[16]{O}. \label{fig:fi}}
\end{figure}

Despite decades of exploration into the \g--process nucleosynthesis in CCSNe, model predictions continue to exhibit discrepancies when compared to solar abundances \citep{roberti:23a}. Specifically, two major challenges persist: (1) the average \g--process yields indicate an underproduction by approximately a factor of three relative to the amount required for explaining the solar abundances of p-nuclei \citep{travaglio:18}, and (2) the \p-isotopes of Mo and Ru systematically experience underproduction by more than an order of magnitude in comparison to other \p-nuclei \citep{rauscher:13,pignatari:16a}. Historically, to estimate the contribution of CCSNe to the solar abundances of \p-nuclei without relying on galactic chemical evolution (GCE) calculations, the \g--process production has been traditionally compared with the production of oxygen. Indeed, massive stars stand as the principal contributors to the galactic production of this element \citep{kobayashi:20}. We therefore define the over-production factor $F_{\rm i}$ for a specific isotope $i$ as the ratio between its mass fraction $X_{\rm i}$ in the supernova ejecta (calculated as the total integrated yield divided by the total ejected mass) and the corresponding solar mass fraction X$_{\rm i,\odot}$ (in our case from \citep{asplund:09}). The average overproduction factor \fz, representing all \p-nuclei, is defined as $F_0 = (\sum_{\rm i=1}^{35} F_{\rm i})/35$. This average factor serves as a proxy for analyzing \p-nuclei production \citep[see, e.g.,][]{arnould:03,pignatari:13}. Since not all the \p-nuclei are solely produced by the \g--process \citep{cumming:85,arlandini:99,goriely:01,dillmann:08,bisterzo:11,bisterzo:15}, Roberti et al.\cite{roberti:23a} defined an additional factor, \fg, as the average over-production of the three most produced \g–only nuclei. To evaluate the \g--process nucleosynthesis in CCSNe, we compare \fz\ and \fg\ with the over-production factor of \isotope[16]{O} (\fo). \figurename~\ref{fig:fi} shows the comparison among $\rm F_i$, \fz, \fg, and \fo, in the case of the same model shown in \figurename~\ref{fig:gammap}. Most of the $\rm F_i$ factors, together with \fz, are under-produced relative to \fo. Instead, \fg\ is a factor of 1.3 higher than \fo. However, given that \p-nuclei are secondary nuclear species, a prerequisite for deeming their synthesis significant is that their over-production must exceed a factor of 2 relative to a primary element such as oxygen \citep{tinsley:80,pignatari:13}.
 
Various alternative nucleosynthetic processes and astrophysical phenomena \citep[see][for a recent overview of the production sites of p–nuclei]{roberti:23a} have been proposed to account for the solar abundances of \p–nuclei. However, the challenge of explaining the distribution of \p–nuclei abundances in the solar composition persists as an open problem. 

\section{Summary and Conclusions}
The evolution, explosion, and nucleosynthesis of massive stars is an extremely challenging problem, both in terms of the computational power required to describe it, as well as in terms of the experimental, observational, and theoretical uncertainties that affect every aspect of this fascinating branch of astrophysics. 

The evolution of massive stars is an inherently three-dimensional problem, given the impact that convection (and to a lesser extent also rotation and magnetic fields) has on the life cycle of stars. However, full 3D simulations of the entire life of the star are practically impossible, and therefore only spherically symmetric simulations are feasible. Experimental nuclear reaction rates, mass loss prescriptions, and convective models are the main sources of uncertainty in these models. Additionally, the size of the nuclear burning network can also have a significant impact, especially in the late phases of the evolution. Finally, different algorithms for nuclear burning, hydrodynamics, and convection can also significantly affect the evolution, and assessing the impact and correctness of these numerical methods on the results is far from trivial. Therefore, different stellar evolution codes often predict different evolutionary paths for a given star, even when adopting similar physical prescriptions. Some promising attempts at simulating small parts of late-stage evolution in 3D are currently underway, and they will be able to provide a better insight into how convection operates in different stellar environments.

An important caveat is that the above discussion is related to single-star evolution. However, about 70 \% of massive stars occur in binaries. Depending on the properties of the stars and their orbit, binary interactions can dramatically change the evolutionary paths of stars. Since only spherically symmetric simulations are currently feasible, correctly describing these interactions is extremely hard, but promising efforts are underway \citep{Fragos2023_POSYDON_methods_paper}.

Uncertainties in single-star evolution also indirectly affect the explosion phase. A core-collapse supernova directly depends on the thermodynamic profiles of the star right before collapse. Therefore, in reality, it is not sensitive to quantities such as ZAMS mass or metallicity. This means that, since uncertainties in the stellar evolution change the final structure of a star with a given ZAMS mass and metallicity, they will also indirectly change the outcomes of a supernova. After the core collapses, there are additional uncertainties that can affect the supernova dynamics. The largest ones are arguably the equation of state of nuclear matter and neutrino opacities. Constraining the equation of state of nuclear matter is particularly challenging since direct experiments at such extreme thermodynamic conditions are not feasible in a laboratory. Therefore, several constraints are based on indirect measurements of microscopic quantities as well as astrophysical observations and theoretical calculations. Neutrino opacities, instead, rely on theoretical calculations and approximations. In recent times, self-consistent simulations of the explosion phase in 3D have become feasible and were able to shed light on several multi-dimensional effects (neutrino-driven turbulent convection \textit{in primis}) which play a key role in the explosion. Finally, recent studies based on physically reliable simulations have started to quantitatively analyze the properties of the explosion, and promising efforts towards predicting the outcome of the explosion based on the pre-supernova properties have already been carried out \citep{Ertl2016_explodability,Muller2016_prog_connection,Wang2022_prog_study_ram_pressure,Gogilashvili2022_FEC,Boccioli2023_explodability}. In the near future, large suites of 3D simulations will also be able to shed some light on this paramount explodability problem.

The chemical composition resulting from CCSNe fundamentally relies on the characteristics of the explosion. At present, the computation of explosive nucleosynthesis using extensive nuclear networks (i.e., including thousands of nuclear species and tens of thousands of reaction rates) is almost exclusively based on post-processing the outcomes derived from one-dimensional simulations. Moreover, the separation between the ejecta and the remnant still depends on a mass-cut that is arbitrarily chosen. However, given that CCSNe are intrinsically multi-dimensional phenomena, relying on yields from 1D models introduces uncertainties that remain challenging to quantify, especially within the context of GCE calculations and studies on the enrichment of the interstellar medium. Relying on 1D simulations may not fully capture the intricate dynamics and nuanced processes inherent in the multi-dimensional nature of core-collapse supernovae, prompting concerns about the accuracy and comprehensiveness of the derived chemical composition. This discrepancy underscores the urgent need for advancements in simulation techniques capable of better encapsulating the intrinsic multi-dimensional nature of core-collapse supernovae, thereby enhancing our capacity to predict and interpret resulting chemical abundances. Within this context, several groups exploring the 3D CCSN problem are yielding promising results in simulating the latest stages of massive star evolution and early phases of the explosion, as well as its impact on chemical evolution \citep{rizzuti:22,rizzuti:23,sandoval:21,sieverding:23,Herwig2000}.


The intricate life cycle of a star, from its formation up until its explosive demise, is a fascinating computational problem. The first semi-analytical and computer models in the fifties and sixties started tackling this extremely tough computational challenge that astrophysicists are still facing today. The tremendous progress in our understanding of stellar evolution, explosion, and subsequent nucleosynthesis was enabled by increasingly more sophisticated codes and increasingly faster CPUs (and, more recently, GPUs). The Exascale era is upon us, and faster and more flexible supercomputers have been recently deployed. This will facilitate large-scale, multi-dimensional, multi-physics simulations that will be able to investigate many phenomena that are currently very uncertain: mass loss, the role of magnetic fields and rotation in the late stages of stellar evolution, the explosion of a CCSN from collapse to shock breakout, nucleosynthesis networks coupled directly to hydrodynamic simulations, and many others.

\vspace{6pt} 



\funding{

L.B. was supported in part by the U.S. Department of Energy under Grant No. DE-SC0004658

L.R. thanks the support from the NKFI via K-project 138031, the European Union’s Horizon 2020 research and innovation programme (ChETEC-INFRA -- Project no. 101008324) and the Lend\"ulet Program LP2023-10 of the Hungarian Academy of Sciences.}


\acknowledgments{}

\conflictsofinterest{The authors declare no conflict of interest.} 






\begin{adjustwidth}{-\extralength}{0cm}

\reftitle{References}



\PublishersNote{}
\end{adjustwidth}
\end{document}